\documentclass[aps,pre,preprint,superscriptaddress]{revtex4-1}
\usepackage{amssymb,mathrsfs,amsfonts}
\usepackage{amsmath}
\usepackage{graphicx}
\usepackage{epstopdf}
\usepackage{nicefrac}
\usepackage{listings}
\usepackage{upgreek}
\usepackage{multirow}
\usepackage{color}
\usepackage{diagbox}
\usepackage{textcomp}

\newcommand{\RN}[1]{%
  \textup{\uppercase\expandafter{\romannumeral#1}}
}

\begin{document}
\title{Key to understanding supersonic radiative Marshak waves 
using simple models and advanced simulations}

\author{Avner P. Cohen}
\email{avnerco@gmail.com}
\affiliation{Department of Physics, Nuclear Research Center-Negev, P.O. Box 9001, Beer-Sheva 84190, ISRAEL}
\author{Guy Malamud}
\affiliation{Department of Physics, Nuclear Research Center-Negev, P.O. Box 9001, Beer-Sheva 84190, ISRAEL}\affiliation{Center for Laser Experimental Astrophysical Research, University of Michigan, Ann Arbor, USA.}
\author{Shay I. Heizler}
\email{highzlers@walla.co.il}
\affiliation{Department of Physics, Nuclear Research Center-Negev, P.O. Box 9001, Beer-Sheva 84190, ISRAEL}


\begin{abstract}
This article studies the propagation of supersonic radiative Marshak waves. These waves are radiation dominated, and play an important role in inertial confinement fusion and in astrophysical and laboratory systems. For that reason, this phenomenon has attracted considerable experimental attention in recent decades in several different facilities. The present study integrates the various experimental results published in the literature, demonstrating a common physical base. A new simple semi-analytic model is derived and presented along with advanced radiative hydrodynamic implicit Monte Carlo direct numerical simulations, which explain the experimental results. This study identifies the main physical effects  dominating the experiments, notwithstanding their different apparatuses and different physical regimes. 

\end{abstract}

\maketitle

\normalsize

\section{Introduction} 
\label{sec:intro}

Radiative heat (Marshak) waves play an important role in many high energy density physics phenomena, such as inertial confinement fusion (ICF) and astrophysical and laboratory plasmas~\cite{lindl,rosen}.
In recent decades, several experiments using supersonic Marshak waves propagating through low density foams have been performed and reported. These experiments facilitating high energy lasers typically use hohlraums as a drive energy generator~\cite{Massen,BackPRL,Back2000,Landen,back_china,TWOP,Dopped1,Dopped2,TWOP2,C8H8,Rosen2007,Keiter2008,Moore2013,Moore2015,Guymer,Fryer2016}. Typically, the drive energy in these experiments is transferred in the form of a heat wave into a low-Z foam, coated with a high-Z envelope (e.g. Au).
The radiative waves are radiation dominated and approximately supersonic (i.e. hydrodynamic motion is negligible), and can be described by the Boltzmann equation. Nevertheless, the high-Z walls are optically thick, and thus affect the system through their ablation into the foam. Consequently,  hydrodynamics should be taken into account, in order to model their effect correctly.

The common numerical schemes for radiation transport usually employed in order to solve these problems, the implicit Monte-Carlo (IMC) and methods of discrete ordinates (the $S_N$ method), have been compared and validated with simple exact benchmarks on several occasions. However, the principal goal of these experiments has been to validate the macroscopic models for radiative hydrodynamics against real experiments, as opposed to simple benchmarks~\cite{Moore2015,Fryer2016}. Hence, the theoretical understanding of these systems is still incomplete, also because of uncertainties in the input microscopic  databases for these models, such as opacity and equation-of-states (EOS)~\cite{Guymer,Fryer2016}.

Most of the experiments examined in the present study were analyzed separately (at different levels) using theoretical models and/or simulations. Still missing, nevertheless, is a unified theoretical modeling and understanding of the different class of all the experiments. Accordingly, the main goal of the present study is to gain a comprehensive understanding and modeling of these experiments, allowing the derivation of a common base ground.
In this work we integrate all the different experiments (that possess sufficient data for modeling), in order to significantly increase  understanding of the radiative phenomena at hand. We present a simple semi-analytic model which takes into account the main physical aspects of the problem. This model yields both qualitative and quantitative results. Nevertheless, we use exact 2D Implicit Monte-Carlo (IMC), coupled to hydrodynamics simulations, in order to attain a detailed reconstruction of the experiments. These two building blocks enable a comprehensive understanding of the physical mechanisms dominating this type of experiment. The simple model was in part previously published in~\cite{CohenJCTT}, a paper that included discussion of some of the main physical aspects of the problem. In the present study, the model is fully derived, including all the main physical procedures. Both the model and the exact simulations are examined against all the experimental results. As will be discussed further below, we demonstrate that although the different experiments were carried out with diverse apparatuses, diagnostics, and target fabrication methods, they share several features, and the main physics governing the system is very similar.

\section{The Experiments} 
\label{sec:expdiscription}
\begin{figure}[htbp!]
\centering 
{(a)
\includegraphics[width=.8\linewidth]{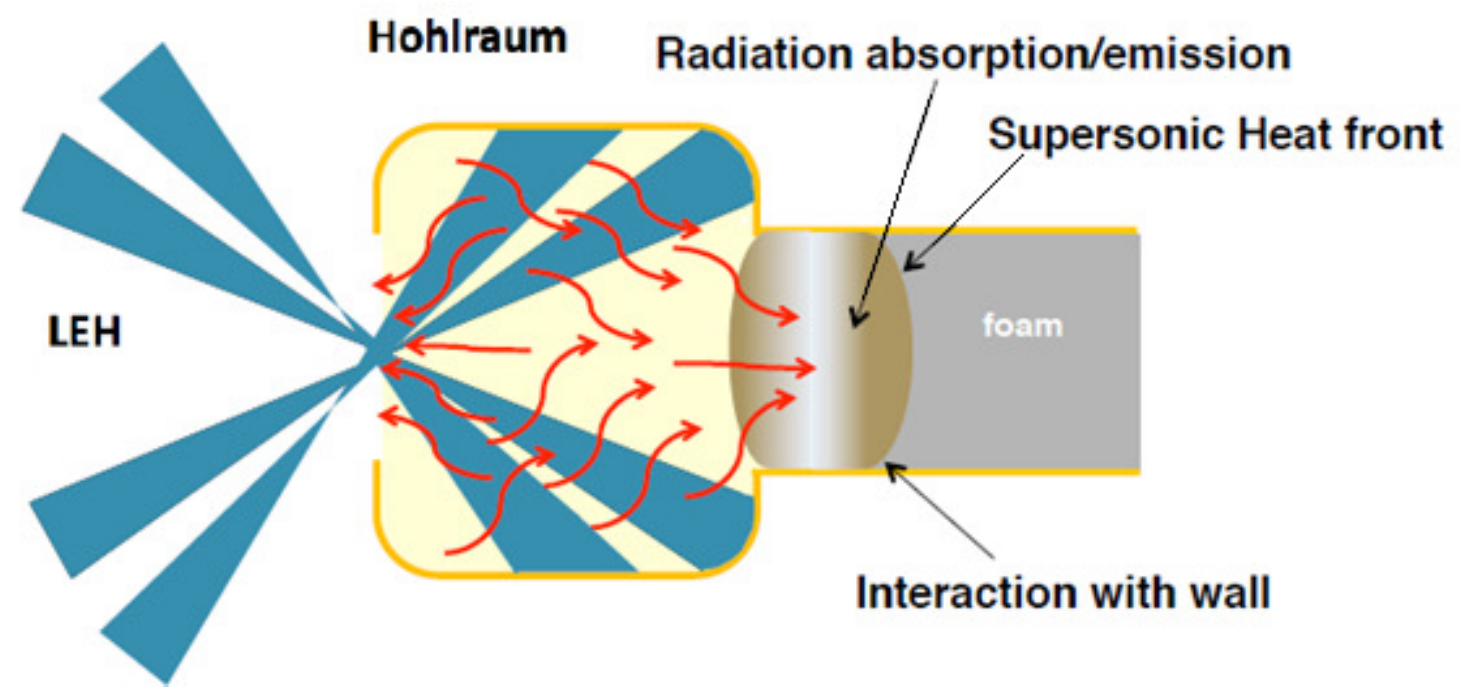}
}
\newline
(b)
\includegraphics[width=.3\linewidth]{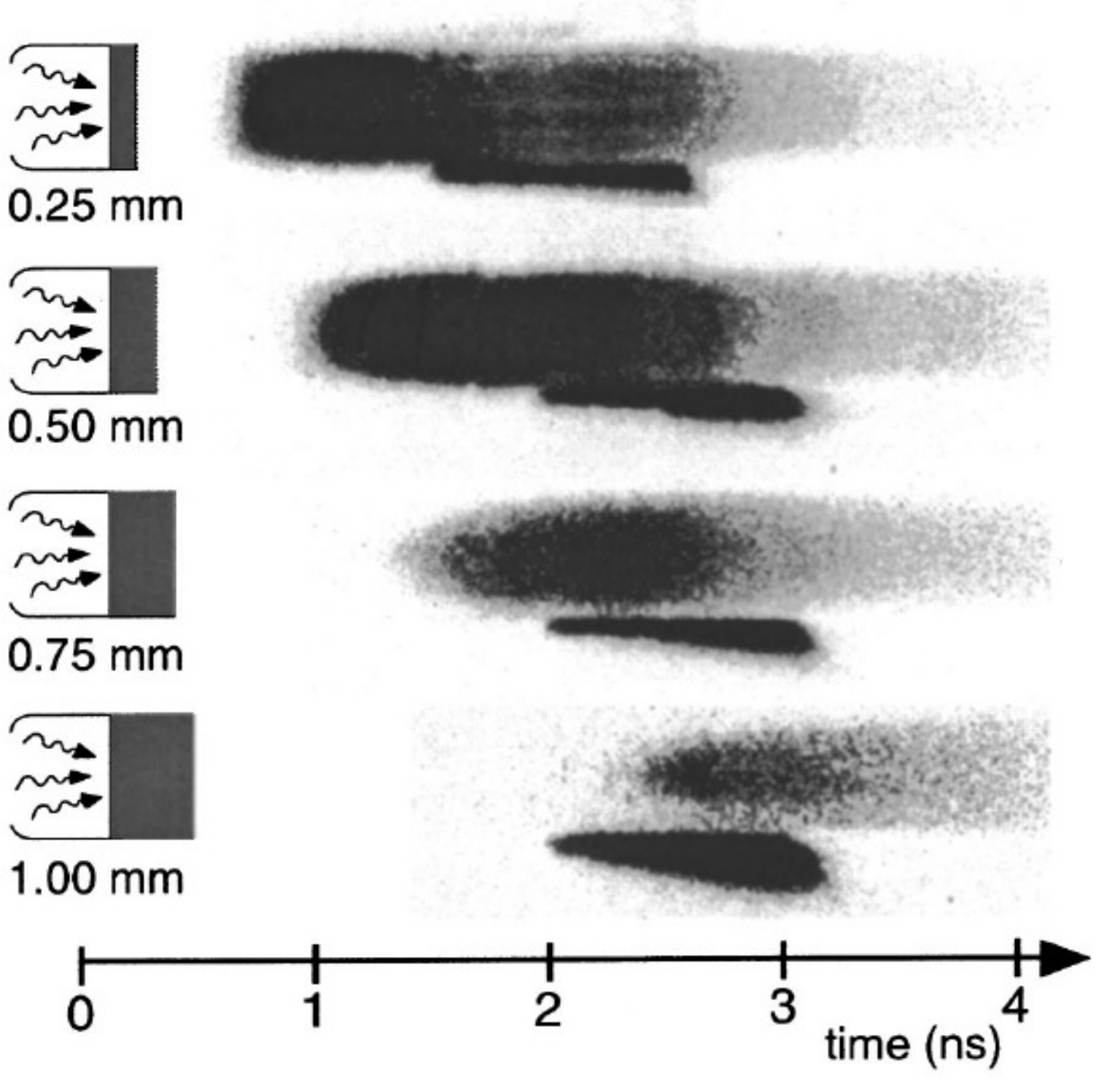}
(c)
\includegraphics[width=.4\linewidth]{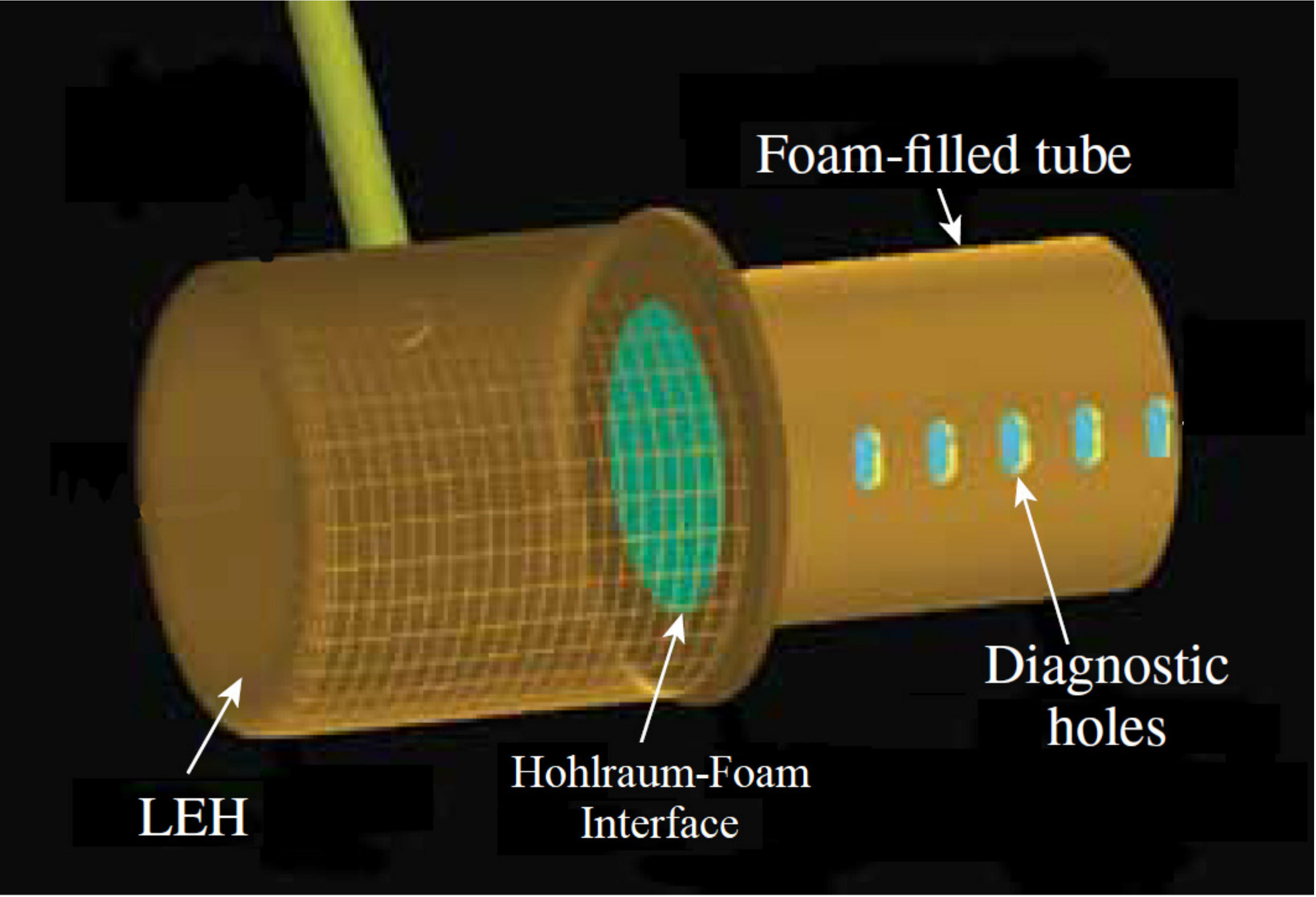}
\caption{Schematic diagrams of typical Marshak waves experiments. (a) The laser beams enter the hohlraum via the Laser Entrance Hole (LEH), and make it function as an X-radiation generator that flows into the low density foam. The latter may be of different lengths in order to track the propagation of the Marshak wave. (b) When tracking the heat wave via different foam lengths, the out-coming flux is measured via XRD or an X-ray steak camera. In the figure, images of the breakout radiation flux from the foam in different lengths by an X-ray streak camera. (c) An alternative experimental technique tracks the heat wave, using a number of slits (holes) in the gold wall, allowing the measurements of the radiation wave emission as it propagates inside the tube. The figures are taken from~\cite{MoorePresentation},~~\cite{Back2000} and~\cite{LLEReview} respectively.}  
\label{fig:schem1}
\end{figure}

During the past decades, several series of experiments have been reported in published literature. The different experiments studied in this paper possess a common procedure, which is presented schematically in Fig.~\ref{fig:schem1}(a).
High energy laser (1kJ-350kJ) is delivered into a small ($\sim$1-3mm), high-Z cavity (usually made of gold), i.e. hohlraum, used as an X-ray source. This shot is represented by the blue beams in Fig.~\ref{fig:schem1}(a). The hohlraum walls absorb the laser energy heated and re-emit soft X-rays, with radiation temperature of about 100-300eV (red arrows in Fig.~\ref{fig:schem1}(a)). A physical package, made of a dilute (low density) foam cylinder is pinned to the hohlraum (through a hole in the hohlraum walls), so that X-rays are delivered into the foam (i.e. the drive energy), generating a heat wave (Marshak wave) that propagates down stream. The foam is usually coated with gold which is optically-thick for X-rays, minimizing possible radiation leakage (the yellow lines around the gray foam in Fig.~\ref{fig:schem1}(a)). Note that since this basic design includes interaction between heat waves and several materials inside the physical package, one must consider the possible hydrodynamic effects of radiation-material interaction. We discuss this further below in Sec.~\ref{sec:Theoretical Background}

Table~\ref{table:1} summarizes the different experiments that have been published in the literature and that are analyzed in this paper. For each experiment we specify the material of the foam, its density, and the maximal drive temperature. Table~\ref{table:1} demonstrates the large range of temperatures, materials (low and high-Z) and densities investigated in this study. Some of the experiments~\cite{Keiter2008,Dopped1,Dopped2} use a small amount of higher-Z doping foams that are compared to ``pure foams". This allows a study of the sensitivity of the Rosseland opacity, (almost) without alteration the heat capacity.
\begin{table}
\begin{center}
\begin{tabular}{||c | c | c | c | c ||} 
\hline
{\bf The experiment} & {\bf Foam Type} & {\bf density} & {\bf Temperature} & {\bf Ref.} 
\\[0.5ex]
 &  & {\bf [mg/cc]} & {\bf [eV]} &  
\\[0.5ex] 
\hline
Massen et al.  &  $\mathrm{C_{11}H_{16}Pb_{0.3852}}$ &  80 & 100-150 & \cite{Massen}   \\
\hline
Back et al.  &  $\mathrm{SiO_{2}}$, $\mathrm{Ta_2O_{5}}$ &  10-50 & 85-190 & \cite{BackPRL,Back2000,Landen,back_china} \\
\hline
Xu et al.  &  $\mathrm{C_6H_{12}}$, &  50 & 160 & \cite{TWOP,Dopped1,Dopped2,TWOP2}   \\
&  $\mathrm{C_6H_{12}Cu_{0.394}}$ &  &  &   \\
\hline
Ji-Yan et al.  &  $\mathrm{C_8H_{8}}$ &  160 & 175 & \cite{C8H8}   \\
\hline
Rosen \& Keiter et al.  &  $\mathrm{C_{15}H_{20}O_{6}}$, &  45-70 & 200-210 & \cite{Rosen2007,Keiter2008}   \\
&  $\mathrm{C_{15}H_{20}O_{6}Au_{0.172}}$ &  &  &   \\
\hline
Moore et al.  &  $\mathrm{SiO_{2}}$, $\mathrm{C_8H_{7}Cl}$ &  90-120 & 310 & \cite{Moore2013,Moore2015,Guymer,Fryer2016} \\
\hline
\end{tabular}
\end{center}
\caption{The different experiments studied in this paper. For each experiment, we specify the material of the foam, its density, and the maximal drive temperature at the entrance to the physical package. For convenience, the references for each of the experiments are given as well.}
\label{table:1}
\end{table}
  
In these experiments, different techniques are employed to examine the heat wave propagation in time. The most popular method is to measure the flux breaking through the edge of the foam as a function of time using an X-ray steak camera~\cite{Massen,BackPRL,Back2000,Landen,TWOP,Dopped1,Dopped2,TWOP2} or an X-ray diode (XRD) array~\cite{Moore2013,Moore2015}, using different foam lengths. An example of measuring the radiative flux using an X-ray streak camera is presented in Fig.~\ref{fig:schem1}(b) taken from~\cite{Back2000}.

Another diagnostic tool used has been to measure the heat wave radiation perpendicular to the heat wave propagation, through a set of small slits (holes) in the gold tubes (see Fig.~\ref{fig:schem1}(c)), tracking the heat wave Eulerian position~\cite{Moore2013,Moore2015}. Another version of this diagnostics technique is to use one long window~\cite{Rosen2007,Keiter2008}. Alternatively the self emission of the foam can be tracked by a back-lighter, which was the technique employed for example in~\cite{C8H8}. However, in this technique the foam should be bare, allowing the radiation of the back-lighter to pass through the material.

\begin{figure}[htbp!]
\centering 
\includegraphics*[width=7.5cm]{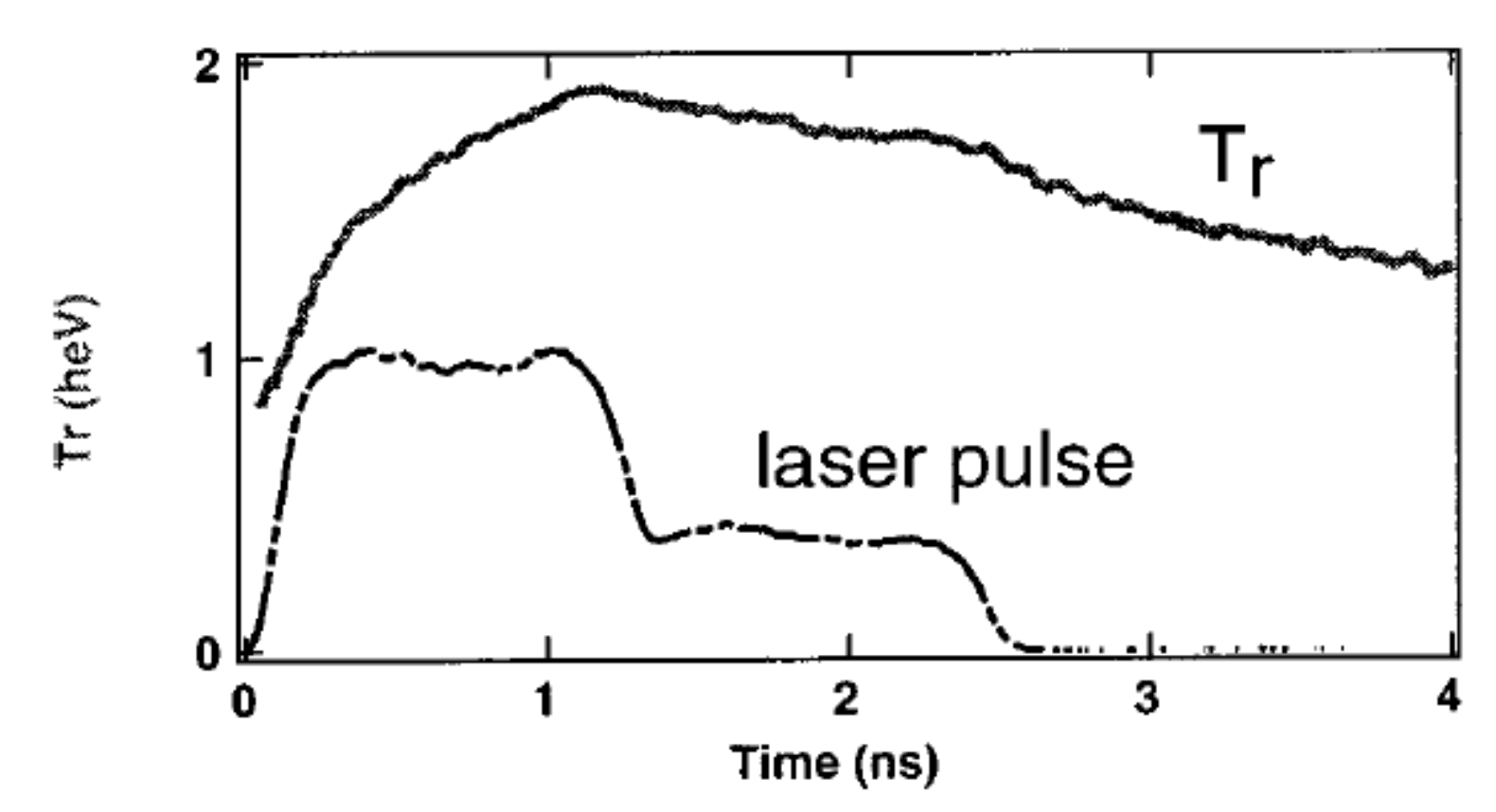}
\caption[BackTd]{A typical example of a measured hohlraum (drive) radiation temperature $T_D(t)$ as a function of the time, measured in the Back high energy experiment. The dash line is the normalized laser pulse shape. The figure is taken from~\cite{Back2000}.}  
\label{fig:BackTd}
\end{figure}

In the different experiments, the radiation drive temperature in the hohlraum is measured as a function of time, which allows an estimation of the incoming temporal flux into the foam. Fig.~\ref{fig:BackTd} shows a typical example of the radiation drive temperature ($T_D$), of the high energy Back et al. experiment~\cite{Back2000}. It should be noted that the radiative temperature in the hohlraum is usually measured via the laser entrance hole (LEH) of the hohlraum (the LEH is shown in Fig.~\ref{fig:schem1}(a)). Interpolation to the exact drive temperature that enters into the foam is not trivial~\cite{Moore2015,MordiLec,MordiPoster}. In this paper we assume that the temperature which was measured via the LEH is equal to the exact drive temperature ($T_D$). 

\section{Theoretical Background} 
\label{sec:Theoretical Background}
The governing equation that describes the behavior of radiative heat waves is the radiative transport equation (RTE), also known as the Boltzmann equation (for photons)~\cite{Pomraning1973}:
\begin{equation}
\begin{split}
& \frac{1}{c}\frac{\partial I(\nu,\hat{\Omega},\vec{r},t)}{\partial t}+\hat{\Omega}\cdot \vec{\nabla}I(\nu,\hat{\Omega},\vec{r},t)+
\left(\sigma_{a}(T_m(\vec{r},t))+\sigma_{s}(T_m(\vec{r},t))\right)I(\nu,\hat{\Omega},\vec{r},t)=\\
& \sigma_{a}(T_m(\vec{r},t)){B}(T_m(\vec{r},t),\nu)+\frac{\sigma_{s}(T_m(\vec{r},t))}{4\pi}\int_{4\pi}I(\nu,\hat{\Omega},\vec{r},t)d\hat{\Omega}+
S(\nu,\hat{\Omega},\vec{r},t)
\end{split}
\label{Boltz}
\end{equation}
where $I(\nu,\hat{\Omega},\vec{r},t)$ is the specific intensity of the radiation at position $\vec{r}$, propagating in the $\hat{\Omega}$ direction with a frequency $\nu$, at time $t$. $B(T_m(\vec{r},t),\nu)$ is the thermal material energy with a frequency $\nu$, while the material temperature is $T_m(\vec{r},t)$, $c$ is the speed of light and $S(\nu,\hat{\Omega},\vec{r},t)$ is
an external radiation source. $\sigma_{a}(T_m(\vec{r},t))$ and $\sigma_{s}(T_m(\vec{r},t))$ are the absorption (opacity) and scattering cross-sections, respectively.
For the experiments discussed in this paper the Boltzmann equation should be coupled to the material energy equation:
\begin{equation}
\frac{C_v(T_m(\vec{r},t))}{c}\frac{\partial T_m(\vec{r},t)}{\partial t}=
\sigma_{a}(T_m(\vec{r},t))\left(\frac{1}{c}\int_0^{\infty}\int_{4\pi}{I(\nu,\hat{\Omega},\vec{r},t)d\nu d\hat{\Omega}}-aT_m^4(\vec{r},t)\right)
\label{Matter1}
\end{equation}
where $C_V$ is the heat capacity, and $a$ is the radiation constant. When the heat wave velocity is close enough to the speed of sound inside the material, hydrodynamics cannot be neglected (i.e. the flow becomes subsonic), and the radiation equations should be coupled to the hydrodynamic equations. In the examined experiments, the heat wave propagating in the foam is supersonic. However, when the gold tube is heated the heat wave within the gold walls become subsonic.

An exact solution for the transport equation is hard to obtain, especially in multi-dimensions. The most well-known exact approaches are the $P_N$ approximation, the $S_N$ method and Monte-Carlo techniques~\cite{Pomraning1973}. In the $P_N$ approximation, we solve a set of moments equations when $I(\nu,\hat{\Omega},\vec{r},t)$ is decomposed into its first $N$ moments. The $S_N$ method solves the transport equation in $N$ discrete ordinates. These two approaches yield an exact solution of Eq.~\ref{Boltz} when $N\to\infty$.
Alternatively, a statistically implicit Monte Carlo (MC) approach can be used~\cite{IMC}. It is also exact when the number of histories goes to infinity.
In the present work, the radiative transfer in the different experiments is modeled via a full
2D IMC model, coupled to the hydrodynamics equations. We now turn to describing the numerical simulations.

\section{2D Simulations} 
\label{sec:Simulations}

\begin{figure}[htbp!]
\centering 
(a)
\includegraphics[width=.49\linewidth]{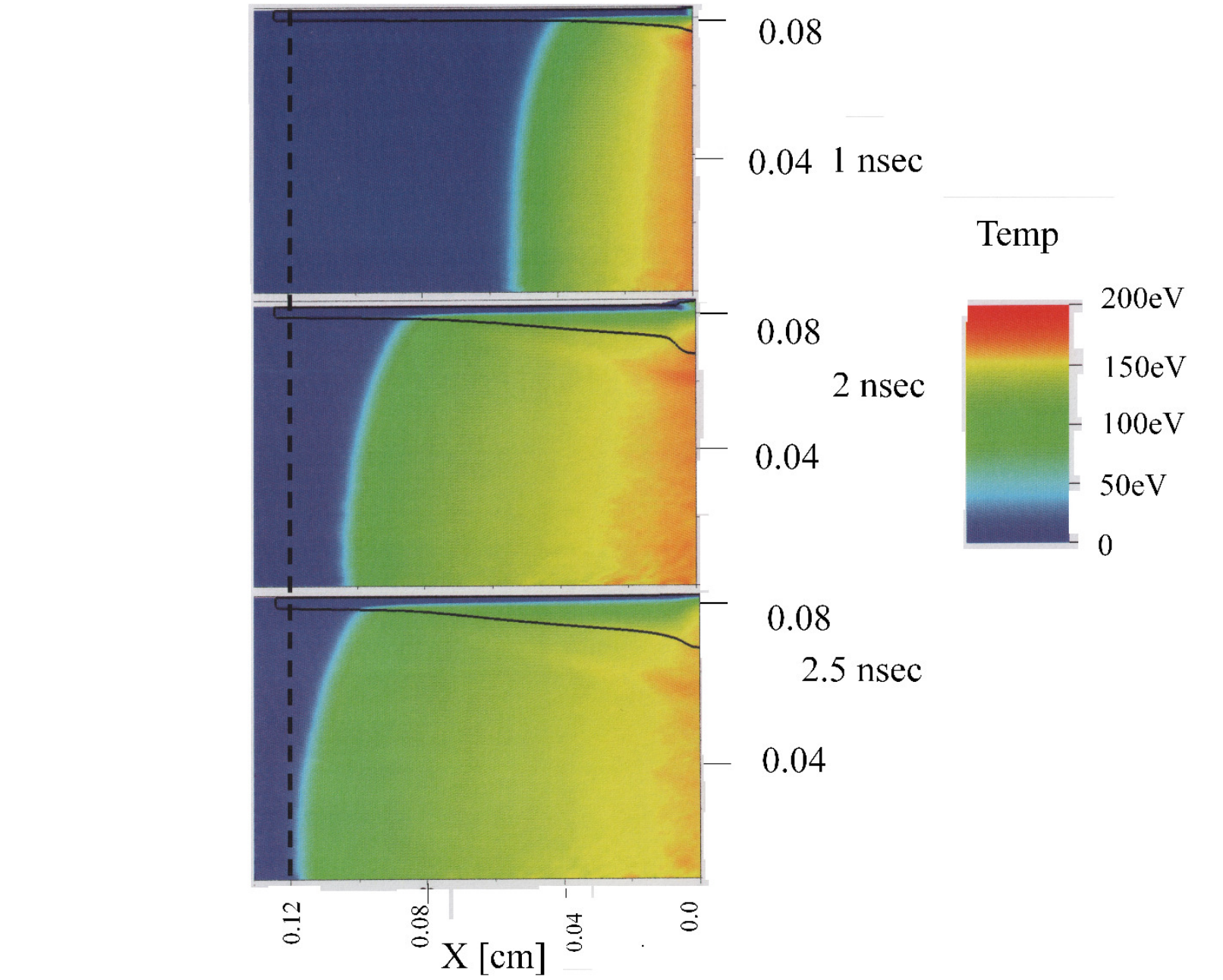}
(b)
\includegraphics[width=.42\linewidth]{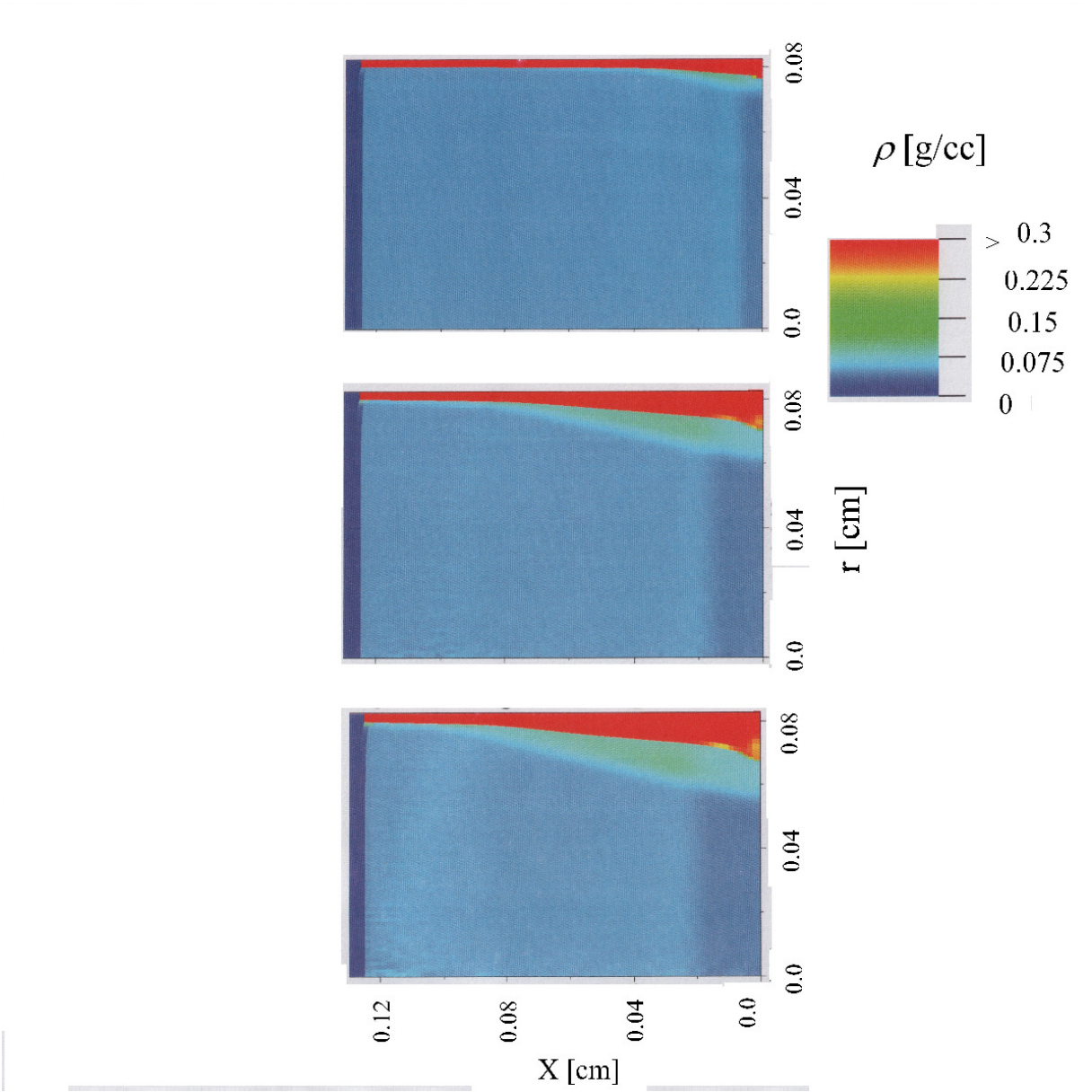}
\caption{The heat wave temperature maps (a) and density maps (b) from a full 2D IMC simulations at different times, $t=1,\, 2,\, 2.5$ \, nsec, in the high energy $\mathrm{SiO_2}$ Back experiment~\cite{Back2000}. At late times, the difference between the heat front in the center ($r=0$) of the sample, and the heat wave that propagates in the side of the sample, can be seen. As the heat wave propagates in $x$'s direction, the ablated gold walls moves into the foam and compressing it.}  
\label{fig:BackMap}
\end{figure}

\begin{figure}[htbp!]
\centering 
(a)
\includegraphics[width=.43\linewidth]{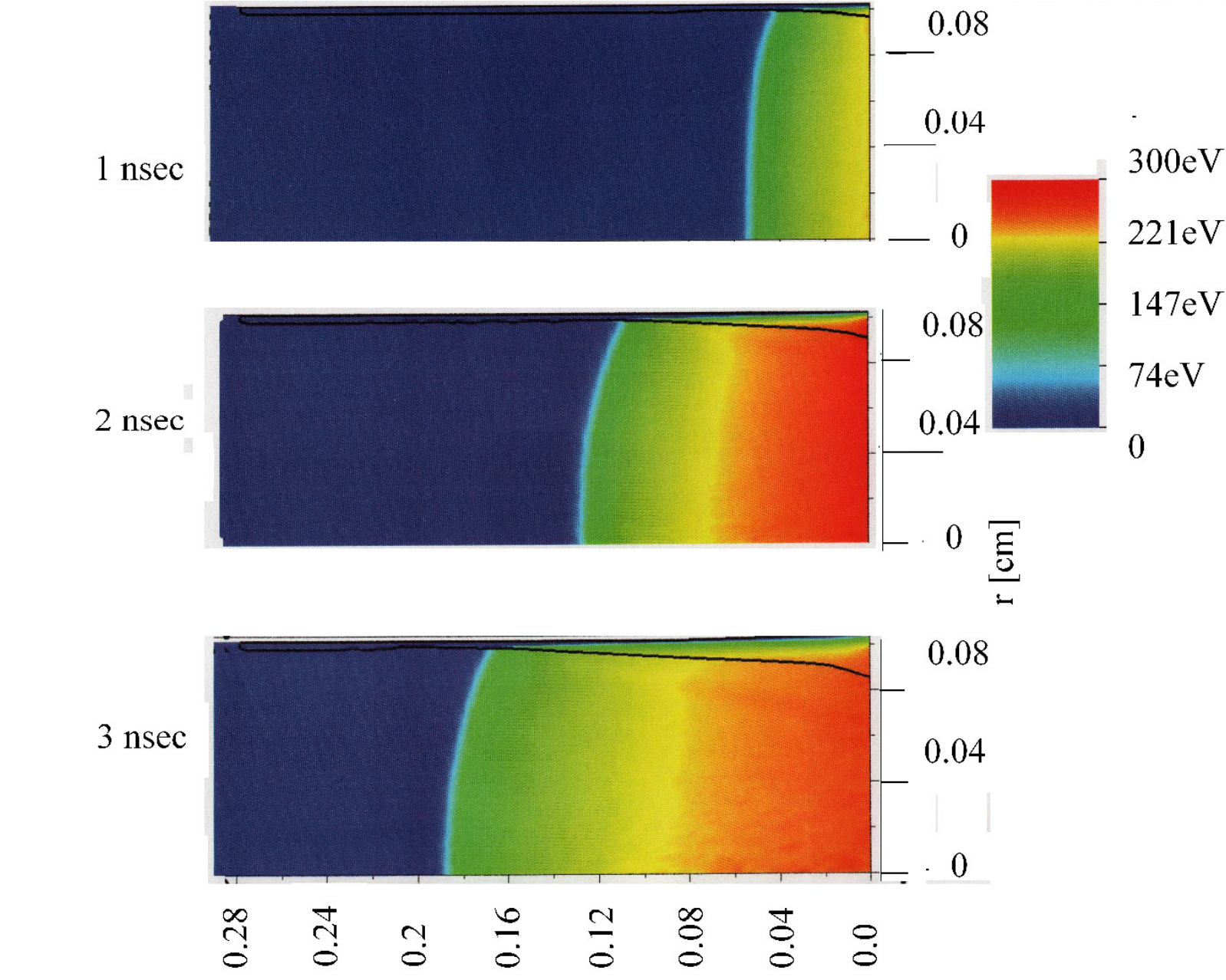}
(b)
\includegraphics[width=.45\linewidth]{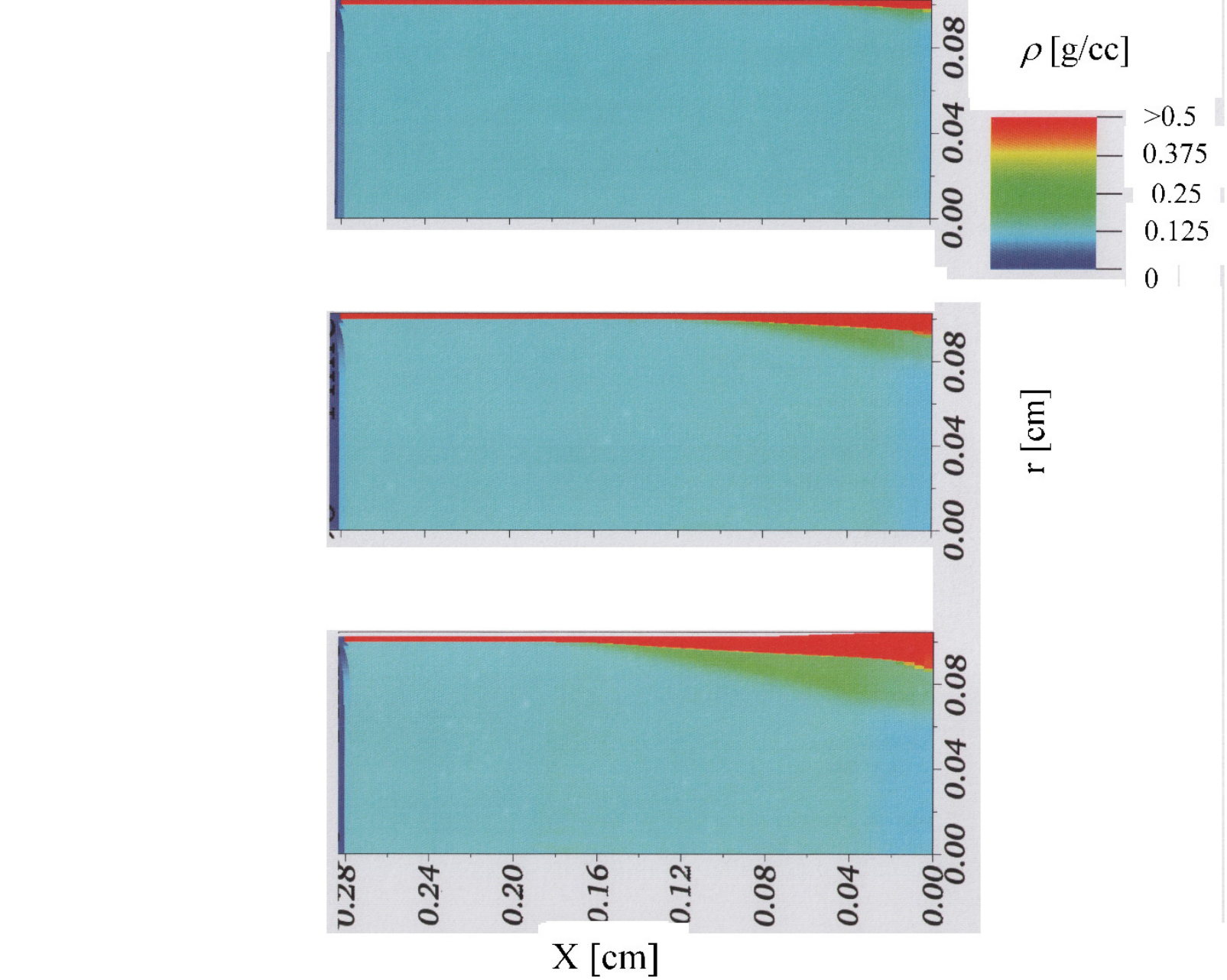}
\caption{The heat wave temperature maps (a) and density maps (b) from a full 2D IMC simulations at different times $t=1,\, 2,\, 3 \, $ nsec, in the high energy $\mathrm{SiO_2}$ Moore experiment~\cite{Moore2015}. }  
\label{fig:MooreMap}
\end{figure}

This section describes the full 2D simulations used for the present study. Demonstration of a 2D radiative hydrodynamics simulation of a propagating heat wave for the high energy Back et al. experiment~\cite{Back2000} can be seen in Fig.~\ref{fig:BackMap} (temperature maps (a) and density maps (b) in different times). Fig.~\ref{fig:MooreMap} shows similar maps of the $\mathrm{SiO_2}$ experiment conducted by Moore et al.~\cite{Moore2015}.

In both examples, the two-dimensional effects can be seen clearly at late times (especially in $2.5$ nsec in  Fig.~\ref{fig:BackMap} or $3$ nsec in Fig.~\ref{fig:MooreMap}), as the heat front is bent along the $r$ direction~\cite{Hurricane2006} due to energy loss to the gold walls. Far from the center the sample becomes denser, due to the hydrodynamic ablation of the opaque walls into the foam (the black line in Figs.~\ref{fig:BackMap} and~\ref{fig:MooreMap} shows the boundary between the foam and the gold).  Another effect that can be clearly seen is the reduction in cross section area on the hohlraum side (right side in the images) due to wall lateral movement. 

\begin{figure}[htbp!]
\centering 
\includegraphics[width=7.5cm]{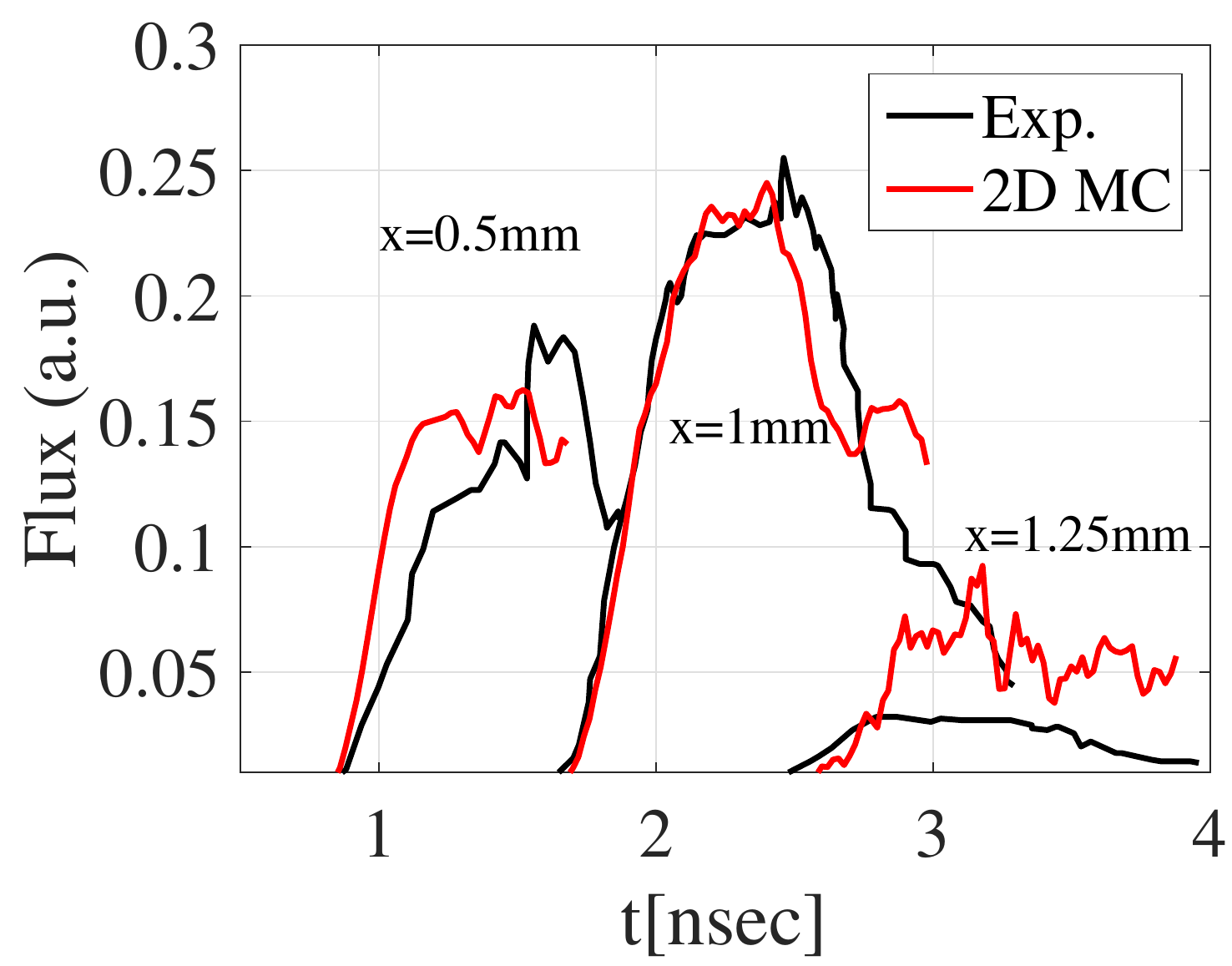}
\caption{
The radiation flux that leaks out from the foam as a function of time for different foam lengths (0.5, 1 and 1.25mm) in the Back et al. high energy $\mathrm{SiO_2}$ experiment~\cite{Back2000}. The experimental results are in black curves, while the 2D simulations are in the red curves.
}  
\label{fig:FLUXBack}
\end{figure}

As discussed above, the most popular diagnostic used in the different experiments was a measurement of the flux that breaks out from the foam as function of time, by an X-ray streak camera~\cite{Massen,BackPRL,Back2000,Landen,TWOP,Dopped1,Dopped2,TWOP2}. Fig.~\ref{fig:FLUXBack} provides an example of a comparison of the experimental measured flux and the current full 2D simulation results for the Back et al. high energy $\mathrm{SiO_2}$ experiment~\cite{Back2000}. A good agreement is achieved, especially in the rising times of the out-coming fluxes. Note that better agreement is evident for shorter foam samples. In this experiment, the breakout time is defined as the time that the intensity is at half max value.

The different experiments that are studied in this work are compared to the simulations by following heat front positions, and in the relevant experiments (when the heat-wave is studied via several foam lengths), the out-coming flux is also examined.

\section{Simple (Semi-Analytic) Model}
\label{sec:analytic}

As already noted, one of the main objectives of the present study is to introduce a simple approximate semi-analytic model, for the purpose of analyzing the experimental results. This model is based on the 1D heat-wave propagation analytic model
of Hammer and Rosen (HR)~\cite{HammerRosen}, while a primary version, that includes only some of the physical phenomena was introduced in~\cite{CohenJCTT}. In the present work, we expand the model, and derive a full version that takes into account 
all the main physical phenomena that affect the general propagation of a radiative heat wave inside a finite tube. We identify the following four separate physical mechanisms that affect the system, each of which is itemized more fully below:
(1) the correct incoming energy flux into the foam; (2) the experimental diagnostics cut-off; (3) the energy loss to wall heating; (4) the effect of wall ablation. It should be noted that the first two mechanisms are one dimensional in nature,
while the last two must consider the two dimensional nature of the problem. Therefore, we separate the discussion relating to 1D aspects/corrections to the model from the 2D effects that have to be taken into account. Using this model we study all
relevant experimental results that were summarized in Table.~\ref{table:1} in Sec.~\ref{sec:expdiscription}. We demonstrate the model using as an example the Back et al. high energy $\mathrm{SiO_2}$ experiment~\cite{Back2000}.
The full comparison with the experimental results is presented below in Sec.~\ref{Analyzing Experiments}. In the next section we present a detailed description of the four physical mechanisms listed above. 

\subsection{1D corrected HR Model}
When the problem contains several mean free paths (mfp), the specific intensity becomes close to isotropic, and the exact Boltzmann solution (Eq.~\ref{Boltz}) tends to a diffusion approximation~\cite{Pomraning1973}. Specifically, the system tends to a local thermodynamic equilibrium (LTE). Hence, both Eqs.~\ref{Boltz} and~\ref{Matter1} can be described by one equation for the temperature of the matter (which is equal to the radiation temperature, under the local thermodynamic equilibrium (LTE) assumption~\cite{Zeldovich2002,Heizler2012}). In this case, the heat wave is characterized by a sharp front, due to the nonlinear behavior of the opacity and the heat capacity.

As a first step before engaging the problem, one must first cover the basic microscopic physical qualities of the material in hand, i.e. opacity and EOS. Numerous studies cover self-similar solutions of both supersonic and subsonic radiative (Marshak) heat-waves~\cite{Marshak1958,Pakula1985,Zeldovich2002,rosenScale2,rosenScale3,HammerRosen,Shussman2015,Malka2016}. In these solutions, one assumes that the Rosseland mean opacity $\kappa$ (which is connected to the absorption cross-section $\sigma_{a}(T_m(\vec{r},t))=(\kappa\rho)$, when $\rho$ is the material's density) and the internal energy $e(T,\rho)$ can be approximated in a power-law form (using~\cite{HammerRosen} notations):
\begin{subequations}
\begin{equation}
\frac{1}{\kappa}=gT^{\alpha}\rho^{-\lambda}
\label{opacityPowerLaw} 
\end{equation}
\begin{equation}
e=fT^{\beta}\rho^{-\mu}
\label{energyPowerLaw} 
\end{equation}
\end{subequations}
In the current work, we are interested in foam parameters (rather then solid gold as in~\cite{HammerRosen}), so $g$, $\alpha$ and $\lambda$ were extracted by fitting Eq.~\ref{opacityPowerLaw} to the opacity spectrum calculated using CRSTA~\cite{Kurz2012,Kurz2013}.  $f$, $\beta$ and $\mu$ were extracted by fitting Eq.~\ref{energyPowerLaw} to the EOS from SESAME tables (when they are supplied)~\cite{SESSAME}, or QEOS~\cite{QEOS} tables in the relevant regime of the experiment (by mean of temperatures and densities). The different parameters for the different material are presented in the Appendix. The parameters for Au were taken from~\cite{HammerRosen}.

\begin{figure}[htbp!]
\centering 
(a)
\includegraphics*[width=7.0cm]{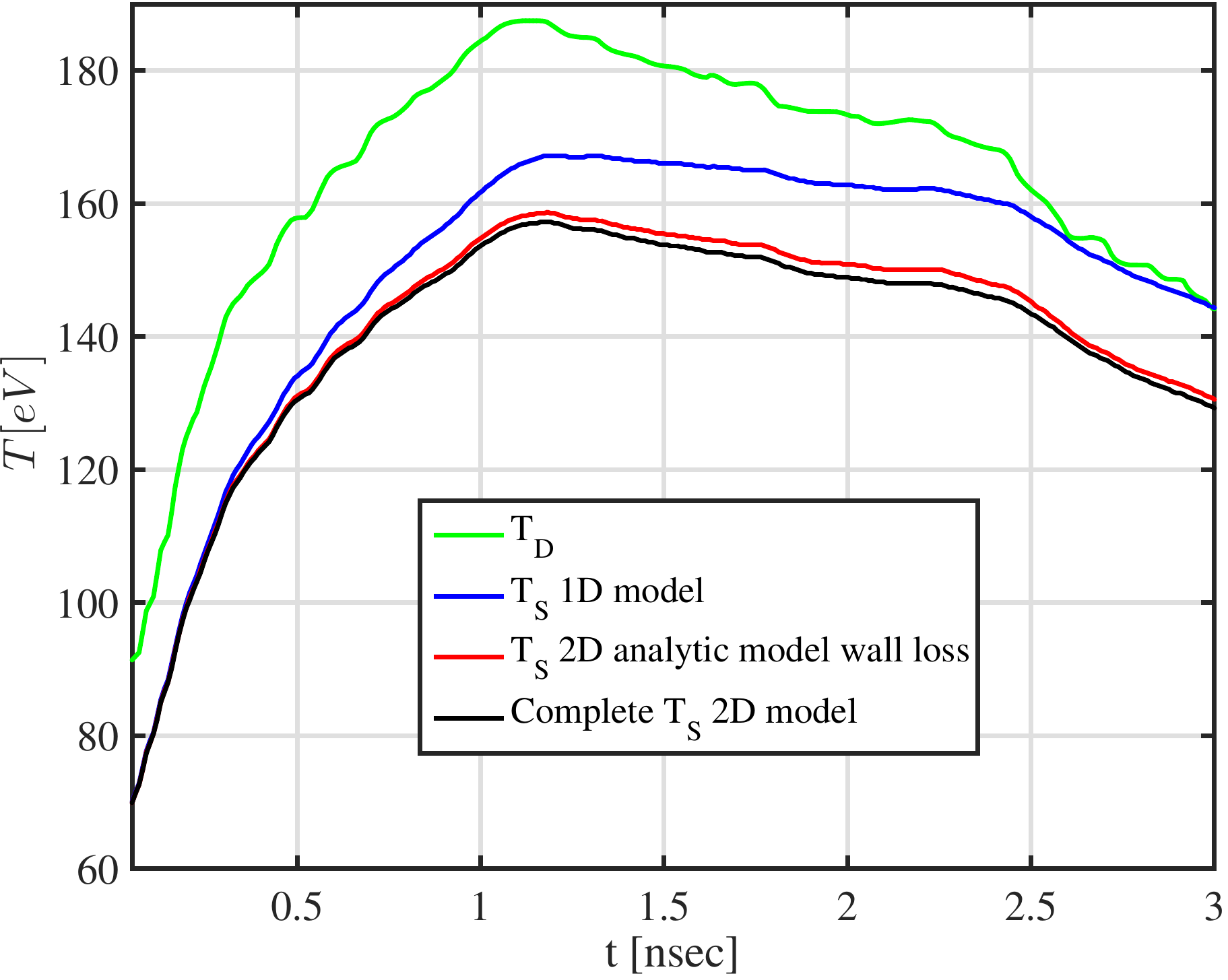}
(b)
\includegraphics*[width=7.3cm]{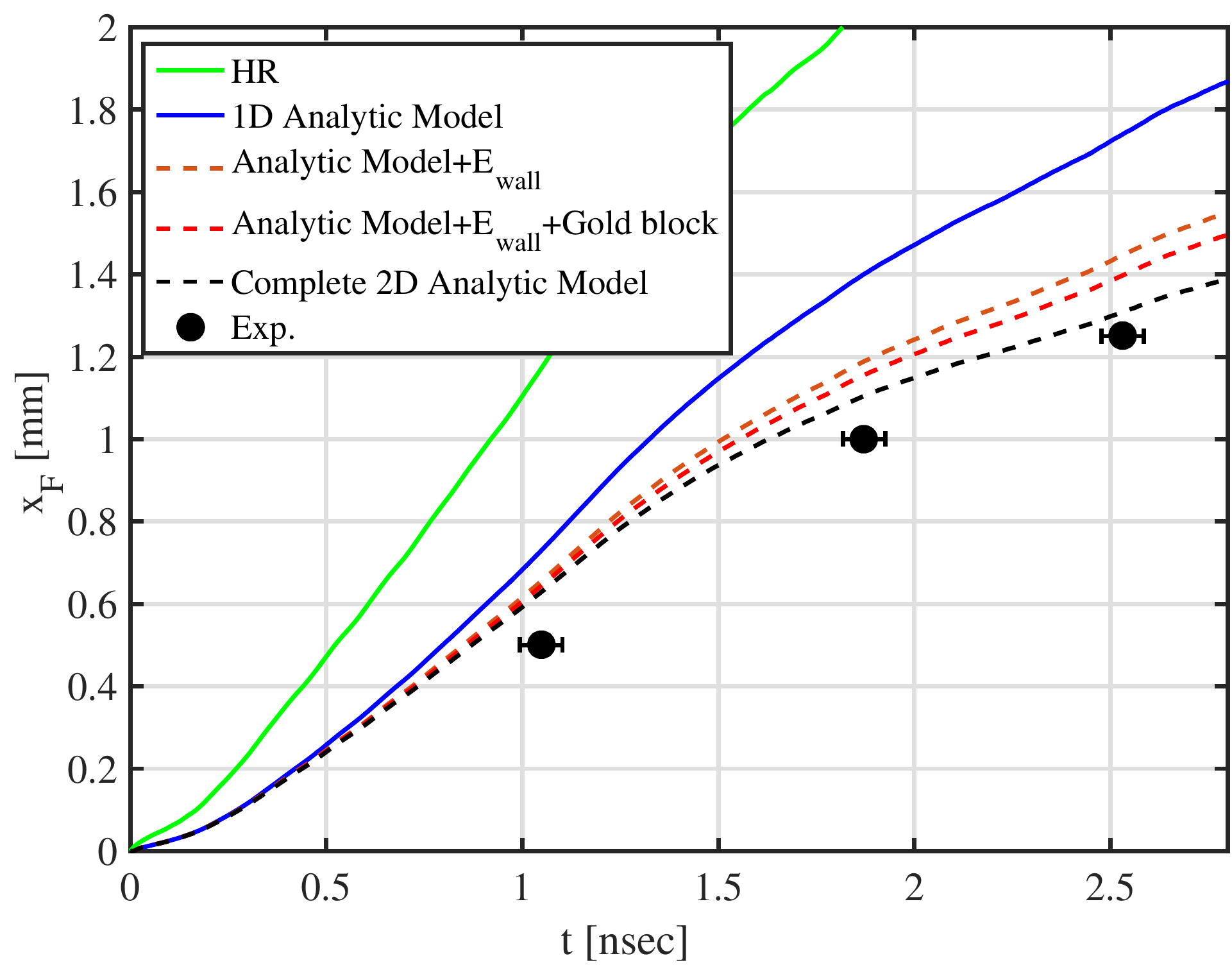}
\caption{(a) The drive temperature $T_D(t)$ (green curve) that was measured in the high energy Back et al. experiment, along with the surface temperatures $T_S(t)$ from the 1D model (blue curve) and the 2D model (red/black curves). The gap between $T_D(t)$ and $T_S(t)$ in the 1D model is due to the re-emitted flux from the foam. The $T_S(t)$ in the 2D model (blue curve) is lower that the 1D, because some of the energy leaks to the gold walls. In the black curve, the 2D temperature is a little bit lower because of the ablation of the gold walls, which blocks some of the incoming flux. (b) The heat wave front $x_F(t)$ as a function of time using the simple model in the Back et al. high energy $\mathrm{SiO_2}$ experiment~\cite{Back2000}. The experimental measurements are marked in circles. The green curve is the 1D {\em naive} HR model. Our 1D modification to HR is introduced in the blue curve. The 2D model which includes energy losses to the gold wall is represented by the orange dashed curve. The 2D model which also includes the gold ablation that blocks part of the energy that enters the foam is in the red dashed curve. The full 2D simple model that includes all the gold ablation effects is in the dashed black curve.}  
\label{fig:XvsTBackPOP}
\end{figure}

Hammer and Rosen (HR) calculated an exact analytic solution for the 1D LTE supersonic diffusion equation using a perturbation expansion theory, for a general surface boundary condition $T_S(t)$~\cite{HammerRosen}. The heat front position, $x_F(t)$, as a function of time is solved analytically and can be expressed as:
\begin{equation}
x_F^2(t)=\frac{2+\varepsilon}{1-\varepsilon}CH^{-\varepsilon}(t)\cdot\int_0^t H(t')dt',
\label{HRXF} 
\end{equation}
where:
\begin{subequations}
\begin{equation}
\varepsilon=\frac{\beta}{4+\alpha}
\end{equation}
\begin{equation}
C=\frac{16}{(4+\alpha)}\frac{g\sigma_\mathrm{SB}}{3f\rho^{2-\mu+\lambda}}
\label{cconstant}
\end{equation}
\begin{equation}
H(t)=T_S^{4+\alpha}(t)
\end{equation}
\end{subequations}
Using the HR solution requires the surface temperature $T_S(t)$ as an input.~A {\em naive} assumption, that the surface temperature is equal to the radiation drive (hohlraum) temperature $T_D(t)$ (the green curve in Fig.~\ref{fig:XvsTBackPOP}(a) for the Back et al. experiments), yields a solution for $x_F(t)$ which is very far from the real experimental results. In Fig.~\ref{fig:XvsTBackPOP}(b) the heat wave front $x_F(t)$ position is presented as a function of time for the $\mathrm{SiO_2}$ Back et al. experiment. It can be seen that taking Eq.~\ref{HRXF} with $T_S(t)=T_D(t)$ (green curve), exhibit with an over estimation of the front velocity, of about a factor of 2 larger than that actually measured. Note that this over-estimation of the heat front velocity is evident for all experiments we consider here. This deviation caused earlier studies presented in the literature to use an {\em ad hoc} factor to decrease the effective $T_S(t)$ to yield an agreement between the theory and the experiments~\cite{
Moore2015,Massen}. Below we consider the physical phenomena which dominates the process of obtaining the correct boundary condition without {\em ad hoc} coefficients or free parameters.

\subsubsection{The different radiation temperatures}

Analyzing the problem, one must distinguish between three different radiation temperatures: The drive (hohlraum) temperature $T_D(t)$, the surface temperature $T_S(t)$ and the brightness temperature of the re-emitted flux, $T_\mathrm{obs}(t)$, that a detector will measure~\cite{MordiLec,MordiPoster}. The latter ($T_\mathrm{obs}(t)$) is the temperature in $\approx1$mfp optical depth ($2/3$ mfp, assuming LTE diffusion behavior)~\cite{MordiPoster}.

Assuming LTE diffusion, the Marshak boundary condition at the surface of the material yields~\cite{MordiLec,MordiPoster,Olson1999,Zeldovich2002,Pomraning1973}:
\begin{equation}
\sigma_\mathrm{SB} T_D^4(t)=\sigma_\mathrm{SB} T_S^4(t)+\frac{F(0,t)}{2}
\label{MarshakBC} 
\end{equation}
Yielding the correct $T_S(t)$ via the given $T_D(t)$ is due to knowing $F(0,t)$, which is the time dependent energy flux on the boundary. HR yields also the total stored energy inside the material~\cite{HammerRosen}, recalling that $F(0,t)\equiv\dot{E}(t)$:
\begin{equation}
E(t)=f\rho^{1-\mu}x_{F}(t)H^{\varepsilon}(t)(1-\varepsilon)
\label{EHR} 
\end{equation}
One can solve Eqs.~\ref{HRXF},~\ref{MarshakBC} and~\ref{EHR} as a closed set of equations, yielding a correct solution for $T_S(t)$. Following~\cite{CohenJCTT}, we use a simple algorithm (detailed in the appendix there) in order to compute $T_S(t)$.

An example of the corrected $T_S(t)$, for the high energy Back et al. experiment is shown in Fig.~\ref{fig:XvsTBackPOP}(a) (blue curve). There is a significant gap between $T_D(t)$ and $T_S(t)$. The complementary $x_F(t)$ for this model is shown in Fig.~\ref{fig:XvsTBackPOP}(b) (the blue curve). 
Using the ``correct" $T_S(t)$ yields a considerable improvement of the HR model (compared to the naive assumption in green curve). However, the model yields $\approx1.3$ faster $x_F(t)$ than the experiment.

\subsubsection{Heat-wave position correction due to experimental cut-off}
\label{henyey_correction}
Another significant 1D feature that needs to be taken into account when comparing the theoretical heat front position, $x_F(t)$, to the experiments is the exact definition of the ``experimental front". This is important especially when measuring the heat flux from the side of the foam (via a hole or bare foam as in ~\cite{Rosen2007,Keiter2008,C8H8} ). The experimental heat front is usually set as a finite value of the maximal emitted flux. For example, in the Keiter et al. experiment the heat front position is set where the radiative energy is $40\%$ of the highest flux~\cite{Keiter2008}, and in the Ji-yan et al. experiment it is $50\%$ of the highest flux~\cite{C8H8}.

In order to calculate the appropriate theoretical estimation, we assume that the temperature profile of the heat wave inside the foam is a Henyey-like profile~\cite{MordiLec,HammerRosen}, with the corrected $x_F(t)$, yields from Eq.~\ref{HRXF}:
\begin{equation}
T_\mathrm{Hy}(x,t)\approx T_S(t)\left(1-\frac{x}{x_F(t)}\right)^{\frac{1}{4+\alpha-\beta}}
   \label{Tshape} 
\end{equation}
In~\cite{CohenJCTT} we show that Eq.~\ref{Tshape} provides a good temperature profile estimation for any given boundary condition with a corrected $x_F(t)$. Assuming that the experimental heat position is determined where the radiative energy is $f$ of the highest flux, the effective energy density of the front is $U'_F(t)=fU_{\mathrm{max}}(t)$ when $U_{\mathrm{max}}(t)=aT^4(0,t)$.
The effective heat front position can be obtained using Eq.~\ref{Tshape}:
\begin{equation}
x'_{F}(t)=x_F(t)(1-f^{(4+\alpha-\beta)/4})
   \label{Henyey} 
\end{equation}
This modification is used in sections~\ref{Keiter_exp} and~\ref{C8H8_exp} to analyze the experiments conducted by Keiter et al.~\cite{Keiter2008}, and Ji-yan et al.~\cite{C8H8}, where the tracking on the heat front was from the side of the foam (see Fig.~\ref{fig:schem1}(c)).

\subsection{2D corrections to the 1D estimation}

Observation of the 2D advance simulations (see Sec.~\ref{sec:Simulations}, Fig.~\ref{fig:BackMap},~\ref{fig:MooreMap}) demonstrates why the heat front propagation slows at late times. This is due to two phenomena. One is the energy loss to the walls (usually gold in most of the experiments) that coat the foam. The other is the ablation of heated walls, blocking part of the energy that enters into the foam  from the hohlraum. The ablation of the walls (when it appears) increases the density of the foam, as shown in Figs.~\ref{fig:BackMap}(b) and~\ref{fig:MooreMap}(b).  We note that one dimensional  simulations, in which no energy is lost to the walls and where the wall ablation is not present, show no such slowing. We also note that similar phenomena were observed in previous hydrodynamic experiments~\cite{Kuranz2018, Trantham2013}.

\begin{figure}[htbp!]
\centering 
\includegraphics*[width=7.5cm]{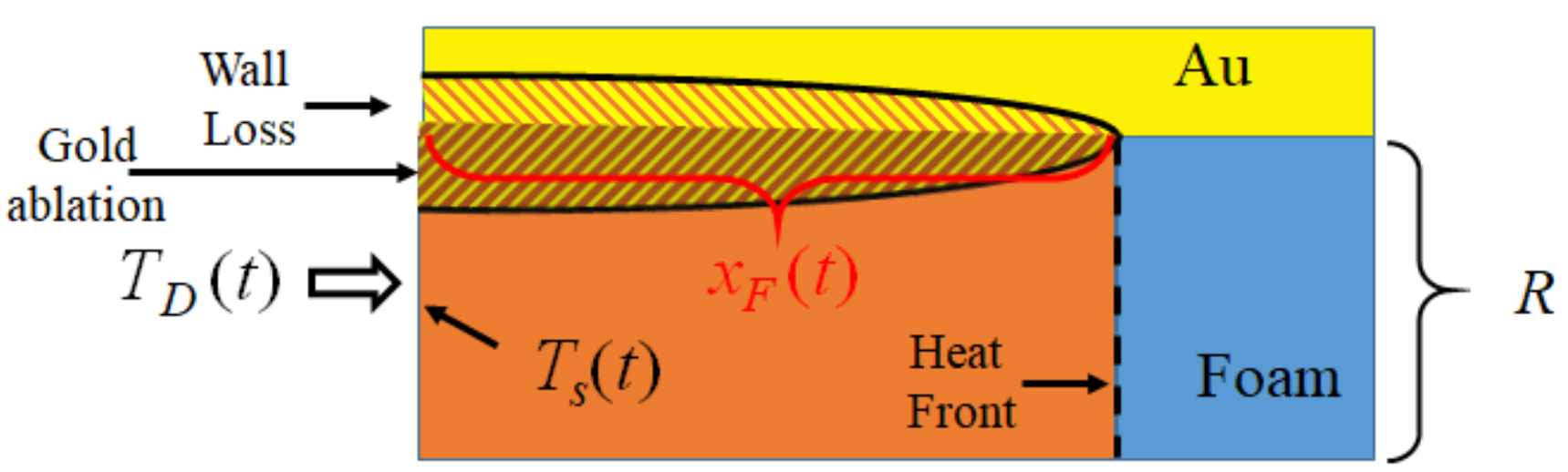}
\caption[WallLoss]{Schematic diagram of 2D slice of the heat wave. The cold foam and the gold walls are in blue and yellow, respectively (the wall area is larger than in reality). The orange area is the heated foam, while the yellow-orange pattern is the heated part of the gold. The heat wave loses energy to the gold walls, and therefore its velocity becomes slower. The walls also ablate into the foam, block part of incoming energy and make the foam denser, and thus slow the heat wave propagation.}  
\label{fig:WallLoss}
\end{figure}

A schematic diagram for this 2D physics is shown in Fig.~\ref{fig:WallLoss}. The cold low density foam and walls are shown in blue and yellow, respectively. The orange area is the heated area inside the foam, when the yellow-orange pattern is the heated area inside the gold walls, that also ablates into the foam. The ablated wall blocks part of the energy that enters the foam due to smaller hohlraum-foam interface area on one hand, and compression in the foam on the other hand, via an ablation shock, laterally propagating inward towards the tube axis. Thus, the heat wave propagation slows due to the density-dependency in Eq.~\ref{cconstant}.
We note that the 2D effect of spatial bending of the Marshak wave due to the wall~\cite{Hurricane2006}, is relatively small, concerning $x_F$, as we will see in IMC simulations later in Sec.~\ref{Moore_exp} (the Moore et al. experiments, Fig.~\ref{fig:MOORE})

\subsubsection{Energy wall losses treatment}

For evaluating the energy losses to the walls, one can use the self-similar solutions of the 1D slab-geometry {\em subsonic} Marshak waves~\cite{Pakula1985,rosenScale2,rosenScale3,HammerRosen,Shussman2015}. Each spatial element is exposed to the heat front in time $t_0(x)$ for a period $t-t_0(x)$, so the total energy that leaks to the wall is the sum of all the energy spatial stored elements (see also~\cite{CohenJCTT}):
\begin{equation}
E_W(t)=2\pi R \int_0^{x_F(t)}\int_{0}^{t}\mathcal{H}(t-t_0(x))\dot{E}_{\mathrm{mat}}(t')dt'dx
\label{EWall} 
\end{equation}
when ${\cal{H}}(t-t_0(x)$ is the Heaviside step function.
We need an expression for $E_{\mathrm{mat}}(t)$, which is the energy in one such element. In most experiments, the walls are made from gold, so we can use the expressions from~\cite{Shussman2015}. Here we use the expression for constant boundary temperature (The dependency of the results to specific BC is small):
\begin{equation}
E_{\mathrm{gold}}(t)=0.59T_0^{3.35}(t-t_0(x))^{0.59}  \quad\mathrm{[hJ/mm^2]}
\label{Egold0} 
\end{equation}

Note that although the general form of  Eq.~\ref{Egold0} is correct, the numerical coefficients should be calibrated for materials other than gold.  For example, in one experiment the foam was coated with a Beryllium (low-Z material) sleeve, exploring the sensitivity to the heat wave propagation to the wall losses (comparing to the high-Z gold walls)~\cite{Back_BE,Landen,Hurricane2006}. The parallel expression for Be is:
\begin{equation}
E_{\mathrm{Be}}(t)=1.27T_0^{4.99}(t-t_0(x))^{0.5}  \quad\mathrm{[hJ/mm^2]}
\label{Egold0be} 
\end{equation}
We note that this expression is for a {\em supersonic} heat-wave (without hydrodynamic motion), since the Be is optically thin, and in $\approx200$eV it is mostly supersonic. As in the foams, the parameters for the  opacity were fitted to CRSTA tables~\cite{Kurz2012,Kurz2013}, and for the EOS, SESAME table~\cite{SESSAME}. The different parameters for the different materials (Gold, Be) are presented in the Appendix.

In some of the experiments, such as the Ji-yan et al. experiment~\cite{C8H8}, the tube possessed no walls, so the leakage can be approximated as an emission to a vacuum~\cite{Olson1999,Zeldovich2002}, instead of Eq.~\ref{EWall}:
\begin{equation}
E_{\mathrm{Leakage}}(t)=2\pi R \int_0^{x_F(t)}\int_{0}^{t}\mathcal{H}(t-t_0(x))\sigma_{\mathrm{SB}}T_0^4(t')dt'dx
 \label{ELeak} 
\end{equation}

We note that considering the 2D effects (as in Eq.~\ref{EWall}), we assume that the heat wave has a flat-top shape inside the foam, i.e. a constant temperature until $x_F(t)$ for simplicity. In~\cite{CohenJCTT} this assumption was checked against a more exact Henyey-like temperature profile, but the effect was small.

In Fig.~\ref{fig:XvsTBackPOP}(a) we can see (red curve) the $T_S(t)$ that is yielded by taking into account the energy losses to the gold walls. We can see the non-negligible difference between the 1D-$T_S(t)$ and the 2D-$T_S(t)$ ($\ap$eV). In Fig.~\ref{fig:XvsTBackPOP}(b) (dashed orange curve) we can see the major improvement achieved by taking into account the 2D effect of energy losses to the walls, that covers about $2/3$ of the difference between the 1D-model prediction and the experimental results.
 
\subsubsection{Ablation of the wall - velocity effects}

\begin{figure}[htbp!]
\centering 
\includegraphics*[width=7.5cm]{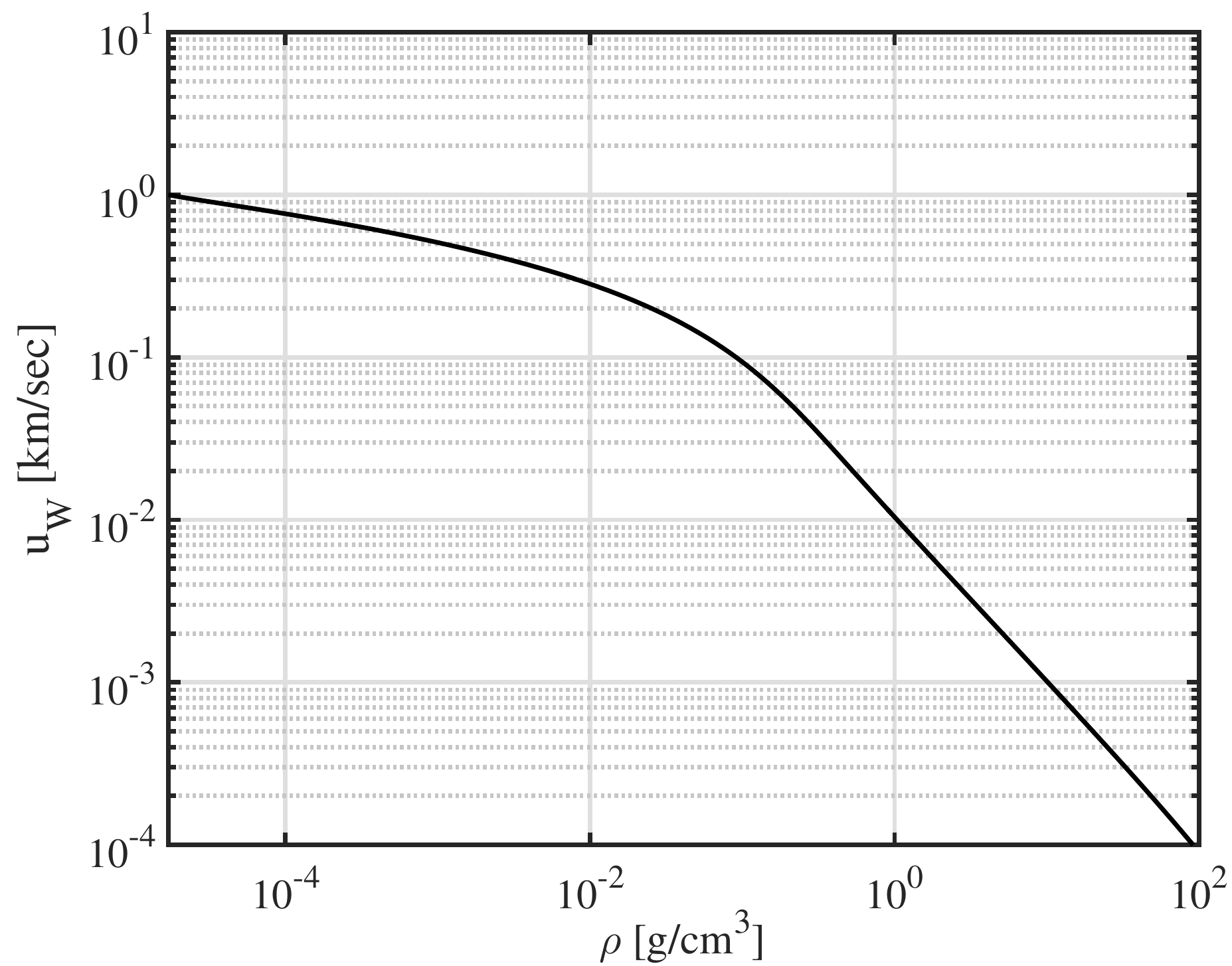}
\caption{The dimensionless velocity multiplier $\tilde{u}$ of gold for Eq.~\ref{Vgold0} as a function of the gold wall density. The parameters are taken from the self-similar subsonic Marshak wave solution for semi-infinite gold~\cite{Shussman2015}.}  
\label{fig:utilde}
\end{figure}

By taking into account the inward lateral ablative motion of the wall one should estimate the rate of ablation. For this purpose, one can use, the subsonic self-similar solution for the ablation velocity. For gold, the ablation velocity is~\cite{Shussman2015}:
\begin{equation}
u_{\mathrm{gold}}(t)=-510.1\tilde{u}(\rho_{\mathrm{gold}})T_0^{0.716}(t-t_0(x))^{0.036} \quad\mathrm{[km/sec]}
   \label{Vgold0} 
\end{equation}
$\tilde{u}(\rho_{\mathrm{gold}})$ is a factor between $\tilde{u}=1$ (at the surface) and $\tilde{u}=0$ (at the heat front inside the gold), which is determined by the self-similar velocity profile. The exact self-similar function as a function of the density is given in Fig.~\ref{fig:utilde}. However, the self-similar solution assumes a free surface (when the density goes to zero), while here the ablative wall is restrained by the finite density foam. Therefore, we should take the value of $\tilde{u}(\rho)$ for the density of the same order of the foam's density. The approximate value is calibrated from the full 2D simulations and found to be $\rho\approx4{\rho_{foam}}$, for the different experiments. 
The coefficients in Eq.~\ref{Vgold0} should be replaced in those cases where the foam is coated with material other than gold. Specifically, in the Be sleeve experiment~\cite{Back_BE,Landen}, we set $u=0$ since the heat wave inside the sleeve is almost supersonic.

Knowing the surface velocity as a function of time, the radius of the foam cylinder for each space interval $x$ at time $t$ is:
\begin{equation}
R(t,x)=R_0-u_{\mathrm{W}}(t)\cdot(t-t_0(x))
\label{radius} 
\end{equation}
As a result, the inlet cross section of the tube decreases over time, decreasing the energy flux from the hohlraum into the foam. The modified entered energy is thus:
\begin{align}
E_{\mathrm{in}}(t)= &\int^{t}_{0}F(t',0)\pi{R^2(t',0)}dt'=\int^{t}_{0}\pi{R^2(t',0)}2\sigma_\mathrm{SB}(T_{D}^4(t')-T_{S}^4(t'))dt'=\\ &2\pi\sigma_\mathrm{SB}\int^{t}_{0}{R_0^2\left(1-\frac{u_{\mathrm{W}}(t')t'}{R_0^2}\right)^2}(T_{D}^4(t')-T_{S}^4(t'))dt' \nonumber
\label{flux} 
\end{align}

In addition to this effect, the foam becomes denser due to the wall ablation: The time-dependent effective mean density of the foam is:
\begin{equation}
\rho(t)=\rho_0\frac{V_0}{V(t)}=\frac{\rho_0V_0}{\int_0^{x_F(t)}\pi R^2(t,x)dx}
\label{rho} 
\end{equation}
Since the heat front propagation velocity depends upon the density in the foam (Eq~\ref{cconstant}), this increase in foam density results in a decrease in the heat front velocity. Eq~\ref{cconstant} can be rewritten as: 
\begin{equation}
C(t)=\frac{16}{(4+\alpha)}\frac{g\sigma_\mathrm{SB}}{3ft}\int_0^{t}\rho^{\mu+\lambda-2}(t')dt'
\label{const} 
\end{equation}

\begin{figure}[htbp!]
\centering 
(a)
\includegraphics[width=7.5cm]{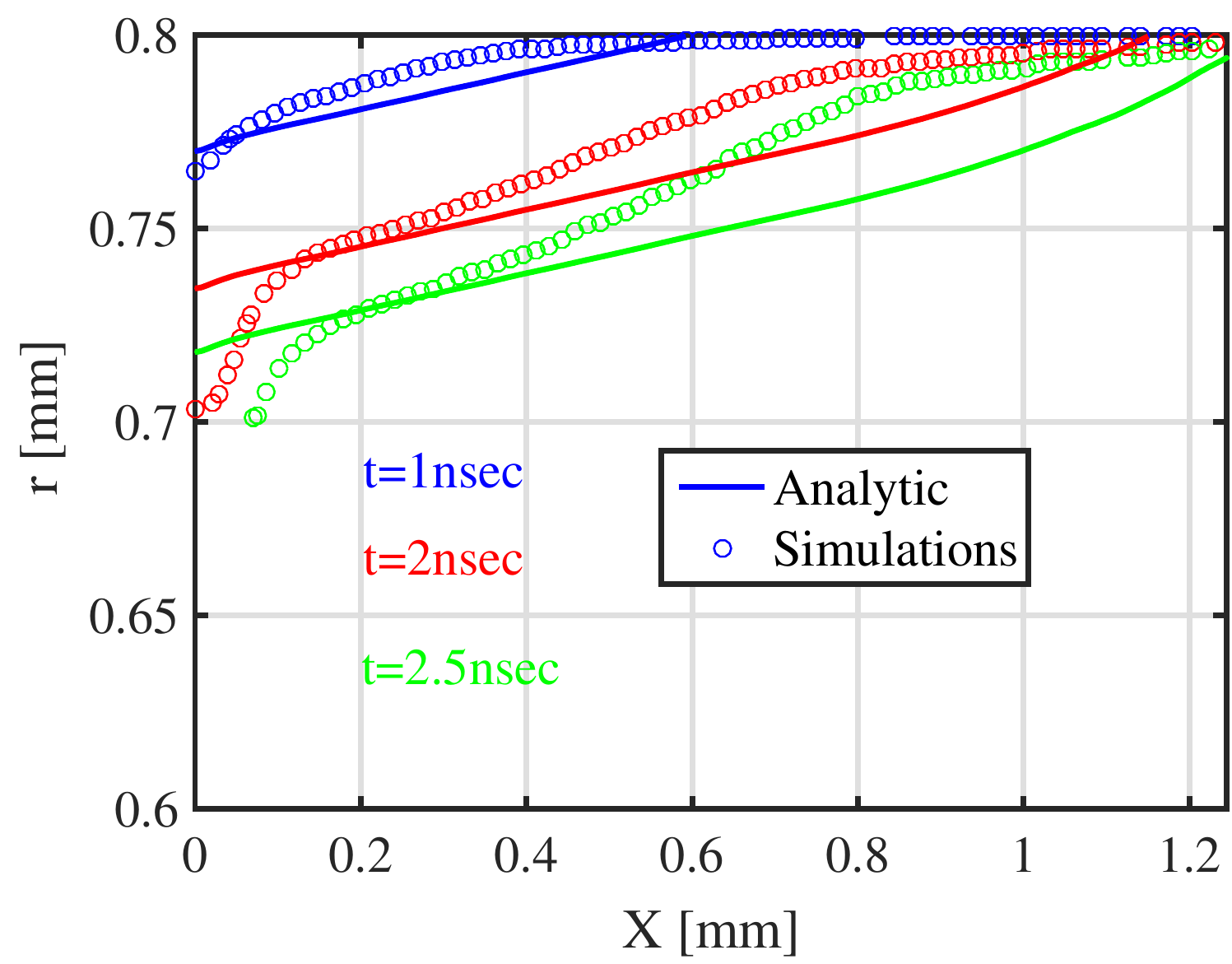}
(b)
\includegraphics[width=7.5cm]{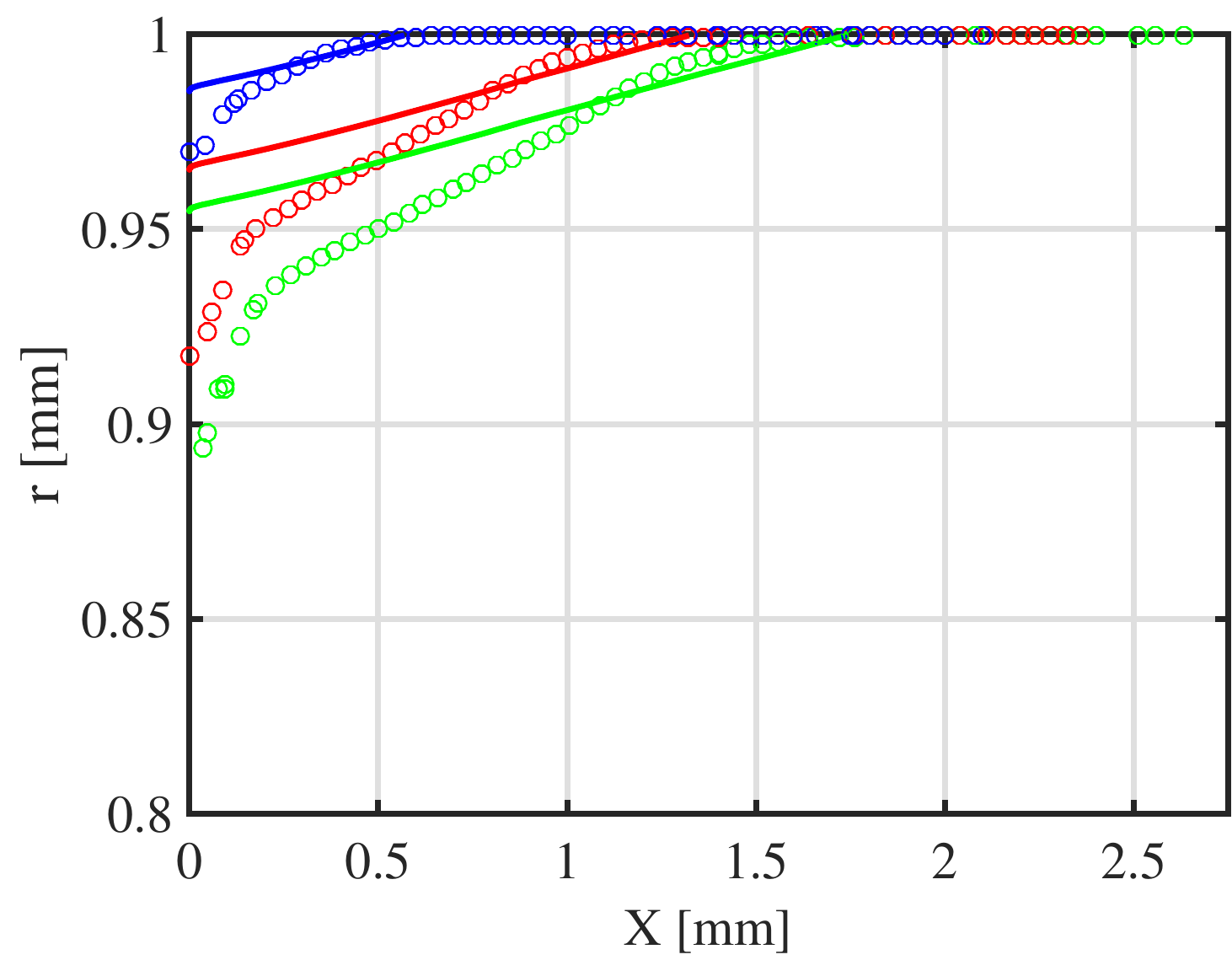}
\caption{Zoom on the interface between the ablated walls and the foam in the high energy Back et al. $\mathrm{SiO_2}$ experiment (a), and the Moore et al. $\mathrm{SiO_2}$ (b). The gold wall is above the lines (larger $r$), and the foam is below the lines. The simulations (circles) and the simple model estimations (solid curves), are presented in $t=1$ nsec (blue), $t=2$ nsec (red) and $t=2.5$ nsec (green).}  
\label{fig:GoldAB2}
\end{figure}

In Fig~\ref{fig:GoldAB2} we compare the ablation front obtained in the simple ``2D Model" against exact 2D IMC simulations for two experiments which have large 2D effects: The high energy Back et al. $\mathrm{SiO_2}$ experiment~\cite{Back2000}, and the Moore et al. $\mathrm{SiO_2}$~\cite{Moore2015}. The comparison shows that the simple model reasonably agrees (up to $\sim 10\%$) with the MC simulations predicted wall position, and therefore captures the wall ablation effect reasonably well.

In Fig.~\ref{fig:XvsTBackPOP}(b) we can see the effects of the ablation on the heat wave front. In the red-dashed curve we can see the effect of the energy-blocking on the foam (which is quite small), while the black-dashed curve is the full 2D model, that takes into account the compression effect of the foam. This yields results that are very close to those in the experimental data (The effect of the ablation of the $T_S(t)$ itself is small, see black vs. blue curves in Fig.~\ref{fig:XvsTBackPOP}(a)). In the next section we examine this model and the 2D simulations in all experimental outlines.
 
\section{Analyzing Experiments}
\label{Analyzing Experiments}
In this section we will review the different experimental measurements, and analyze them by using the simple model along with 2D radiative hydrodynamics simulations. 

\subsection{The Massen et al. experiment}
\label{Massen experiments}

The first reported quantitative measurements of a heat wave propagation were reported by Massen et al.~\cite{Massen}. The experiments were carried out at the GEKKO-XII facility with a maximal energy of $2.5$kJ, delivered in a pulse of 0.8-0.9 nsec duration. The experiments were carried out with different lengths of tubes: $50,{\,}75,{\,}100,{\,}150,{\,}200,{\,}$ and $300{\mu}$m. The foam was $\mathrm{C_{11}H_{16}Pb_{0.3852}}$ (i.e. involving high-Z material) with a density of $80\mathrm{mg/cm^3}$. The study reports three different (constant) drive temperatures $T_D=100$eV, $120$eV and $150$eV, obtained by changing the hohlraum dimensions. The temporal profile of the radiation from a hole covered with the foam, and an open reference hole (fiducial) was measured by a soft X-ray streak camera (XRSC).
\begin{figure}[htbp!]
\centering 
\includegraphics[width=8cm]{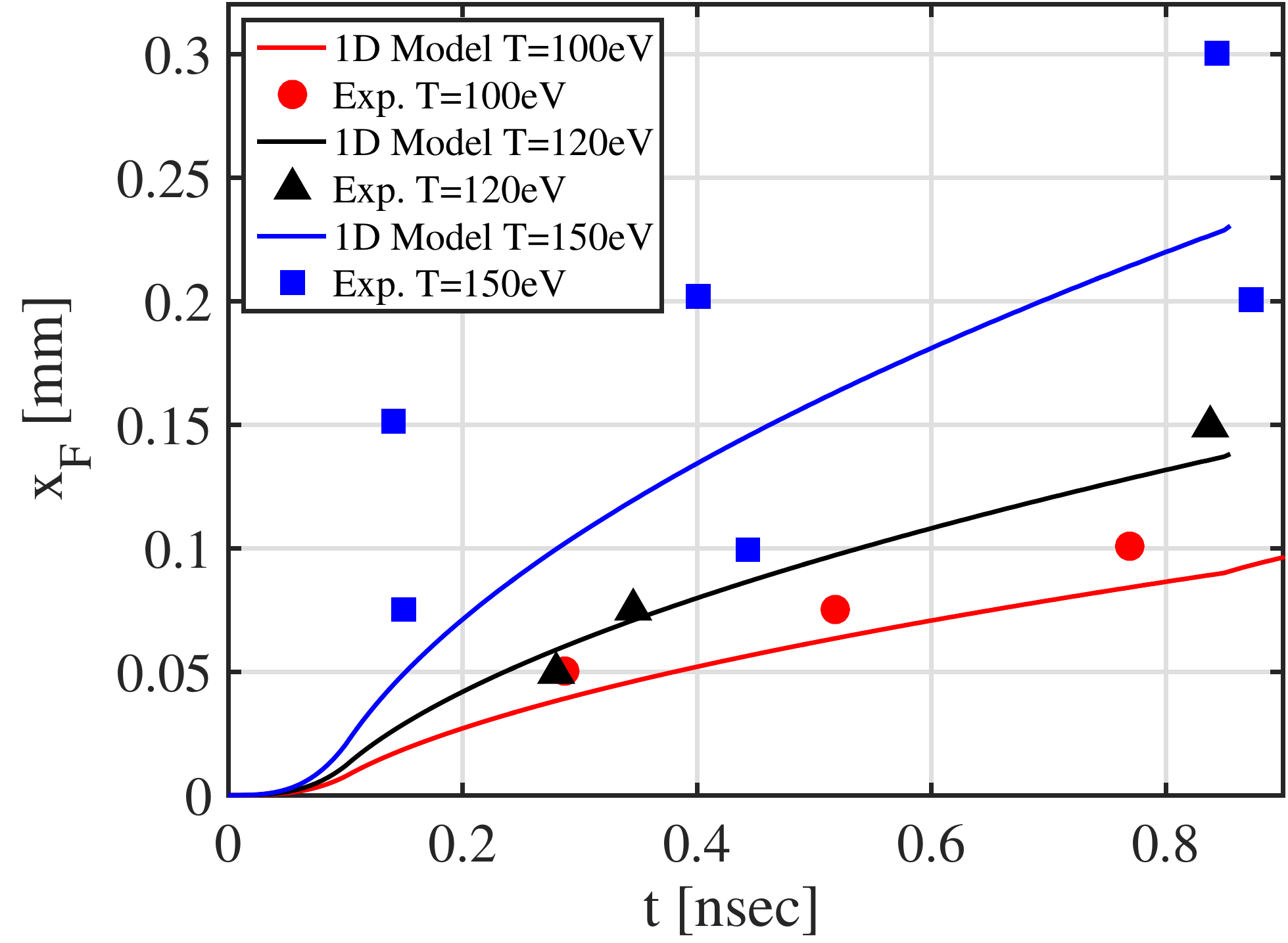}
\caption{A comparison of the simple 1D model predictions (smooth lines) with the Massen et al. experiment~\cite{Massen}. The heat wave front $x_F(t)$ as a function of time for different drive temperature $T_D$, $100$eV in red, $120$eV in black, and $150$eV in blue. The 1D simple model yields results close to the general trends of the experimental data, especially for the low drive temperatures.}  
\label{fig:Massen}
\end{figure}

It can be seen that the 1D simple model (solid curves in Fig.~\ref{fig:Massen}) yields good agreement with the experimental results for the low drive temperatures, while in the high drive ($T_D=150$eV) temperature (blue squares in Fig.~\ref{fig:Massen}), the experimental data has a large spread, preventing a clear comparison. Even so, the basic trends are captured and reproduced by the model. We note that using the original naive HR estimations yields much faster velocities than our corrected model. Deviations of the naive estimations from the experimental data are about $20\%$. For that reason, Massen et al. reported an {\em ad hoc} correction (that was calibrated from a full simulation) for the heat-wave propagation velocity by a factor of $\sim{0.8}$, taking into account the re-emitted flux from the foam, back to the hohlraum~\cite{Massen}. In this work we have presented the physical explanation for this correction, quantitatively.

Since in this experiment the drive temperature was given as constant, and the foam diameter was not reported, we limit the analysis in this experiment to the 1D simple model. We also note that the difference between the simulations and the model are much smaller than the scatter of the experimental data.  

\subsection{Xu et al. experiments}
\label{Xu experiments}
Several studies reported about a decade ago showed a set of heat wave measurements conducted at the SG-II facility using $\mathrm{C_6H_{12}}$ foams~\cite{Dopped1,Dopped2,TWOP,TWOP2}. The radiation drive temperature $T_D$ that was reported and has been used for analysis in the current study is shown below in Fig.~\ref{fig:XvsTTWOP}(b) (green curve). The experiments were carried out in 50mg $\mathrm{C_6H_{12}}$ with different lengths, 300${\mu}$m and 400${\mu}$m~\cite{TWOP}. The main diagnostic that was used to track the heat wave propagation was also an XRSC in several different energy lines. The major line was around 210eV. An additional version of this experiment used a 300${\mu}$m copper doped $\mathrm{C_6H_{12}}$ foam ($\mathrm{C_6H_{12}Cu_{0.394}}$)~\cite{Dopped1,Dopped2}. 
\begin{figure}[htbp!]
\centering 
\includegraphics[width=7.5cm]{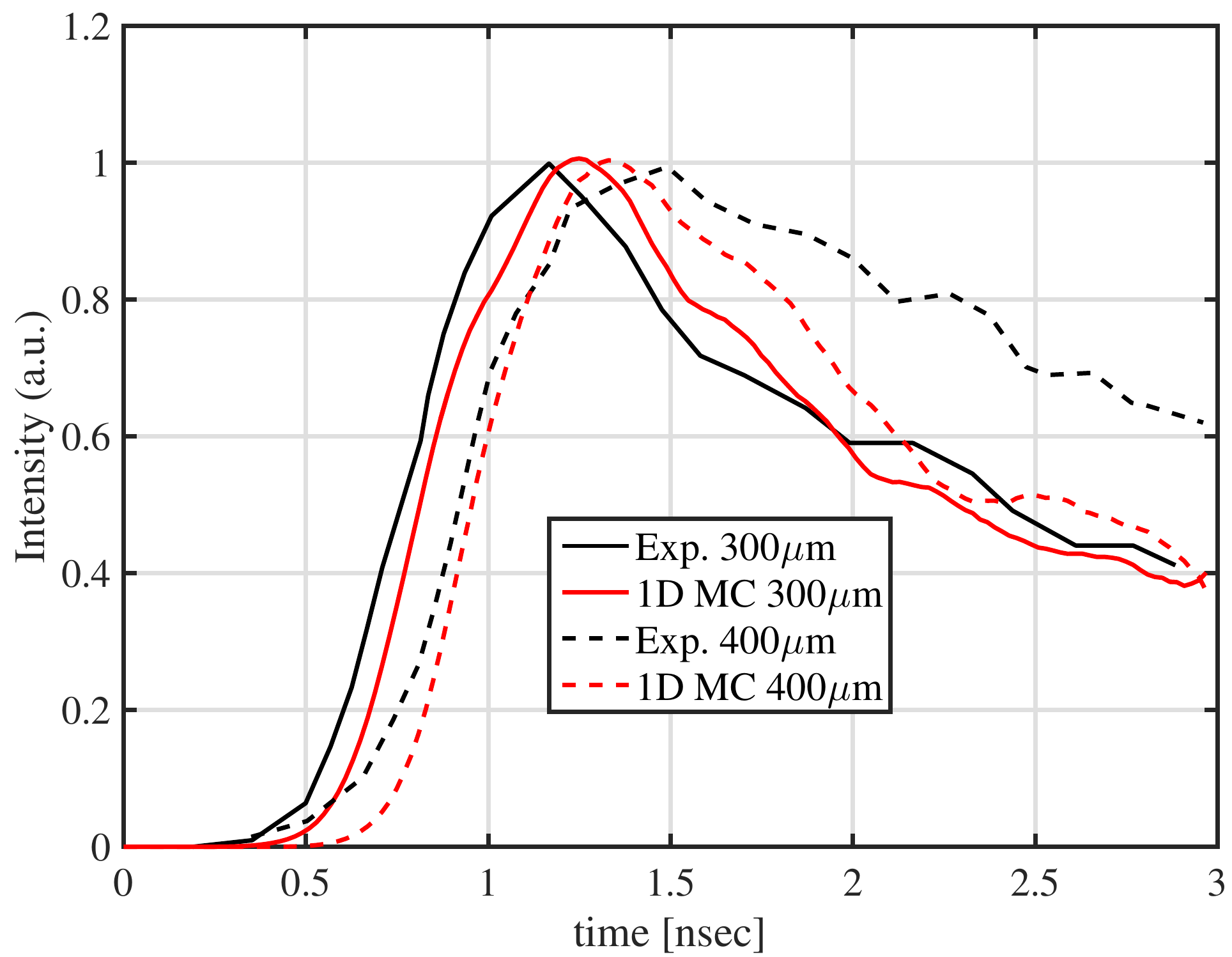}
\caption{The radiation flux that leaks from the edge of the foam as a function of time, in different foam lengths.
The experimental measurements are shown by the black curves and are taken from~\cite{TWOP}. The red curves represent 1D IMC simulations. The simulations and the experiments match in less than 50psec, and the variance between the leaked flux of 300$\mu$m and 400$\mu$m is similar to the experimental variance.}  
\label{fig:FLUXTWOP}
\end{figure}

A comparison of the intensity of the flux that leaks from the edge of the foam  as a function of time, between the experimental data and the 1D IMC numerical simulation is presented in Fig.~\ref{fig:FLUXTWOP} (black curves). The simulations show good agreement with the data, with a difference of less than $50psec$ in breakout times, defined as the time the intensity is half of the maximum intensity. Moreover, the experimental results and the simulations show very similar variance between the $300\mu$m and the $400\mu$m. We note that by the time the heat wave breaks out of the foam, the gold tube around the foam ($600\mu{m}$ diameter) does not heat significantly. Therefore, simulations for this experimental setup demonstrate low sensitivity to 2D or hydrodynamic effects.

The experimental breakout times are given in Fig.~\ref{fig:XvsTTWOP}(a) in black circles. The doped foam ($\mathrm{C_6H_{12}Cu_{0.394}}$) result that was attained with $300\mu$m is the red circle. Using the naive HR solution (were $T_S(t)=T_{D}$), yields an over estimation of $x_F(t)$, with no agreement with the experimental results (blue and green curve for the pure $\mathrm{C_6H_{12}}$ and doped $\mathrm{C_6H_{12}Cu_{0.394}}$ foams, respectively). Results from our 1D models are shown in the solid curves, whilst the full 2D model results are represented by the dashed curves.
\begin{figure}[htbp!]
\centering 
(a)
\includegraphics[width=7cm]{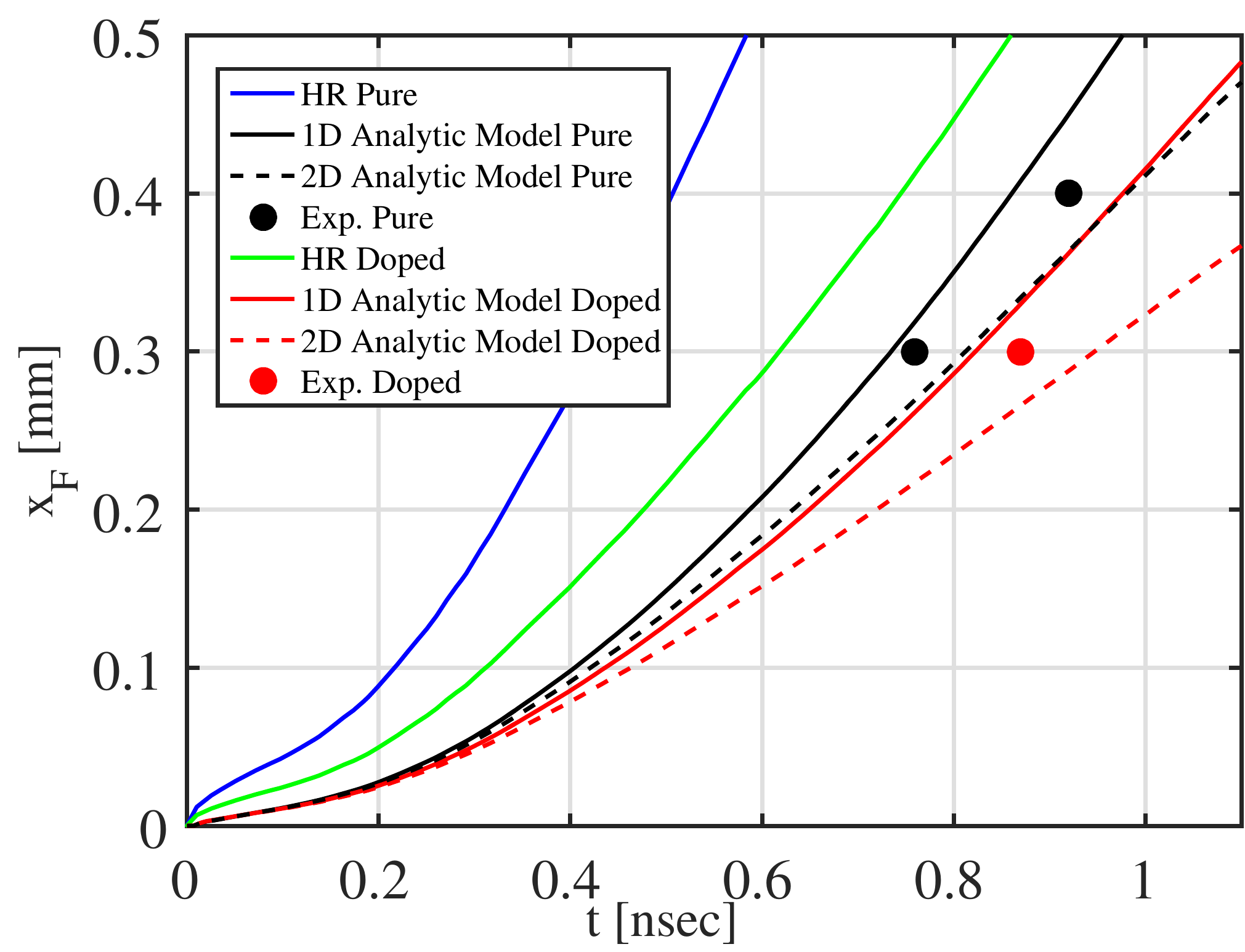}
(b)
\quad\includegraphics[width=7cm]{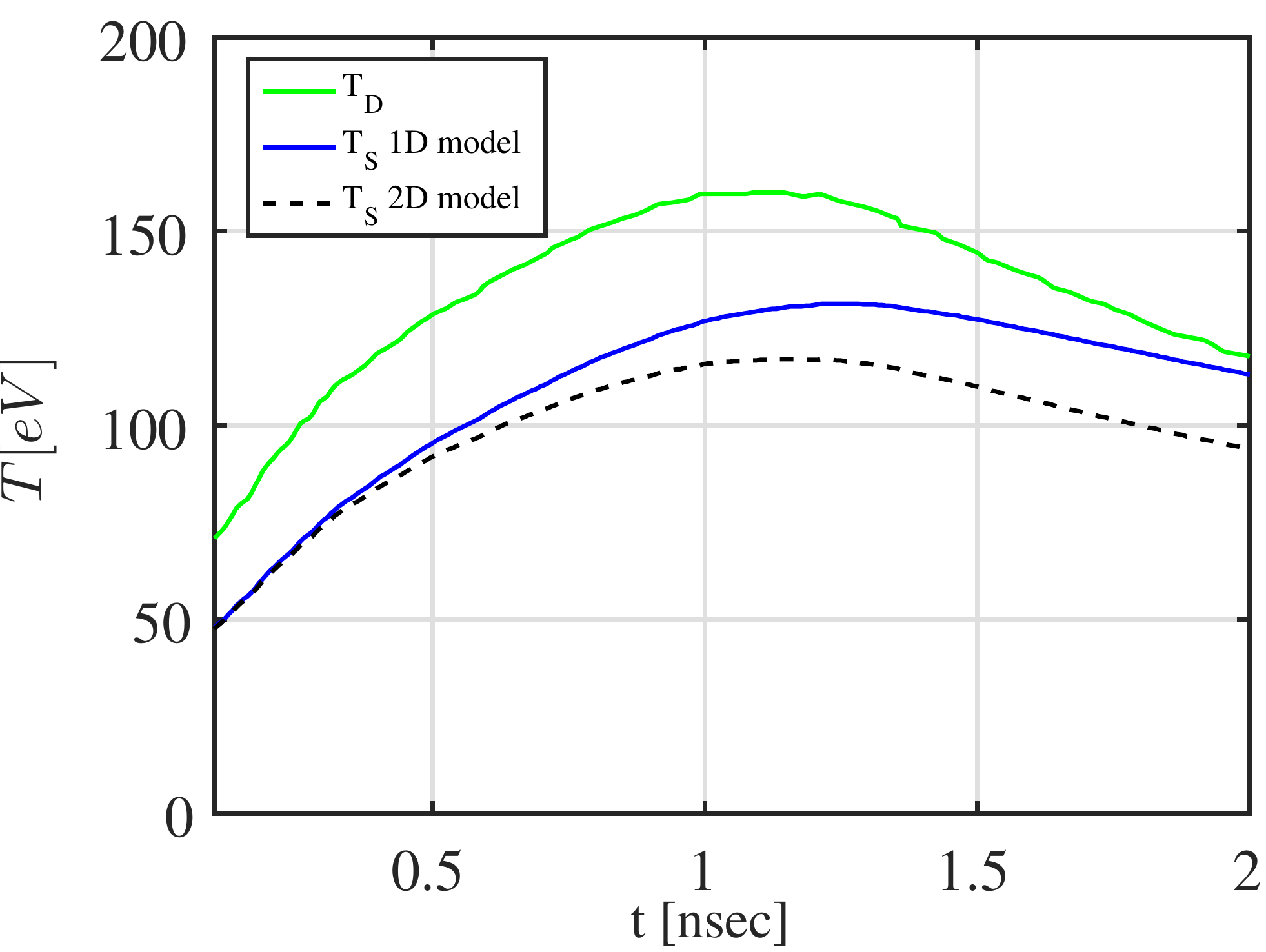}
\caption{(a) The heat wave front $x_F(t)$ as a function of time for different simple models. The experimental data are marked in black circles for the pure $\mathrm{C_6H_{12}}$ and are taken from~\cite{TWOP}, and in the red circle for the doped $\mathrm{C_6H_{12}Cu_{0.394}}$ and taken from~\cite{Dopped1,Dopped2}. The blue and green curves are the 1D {\em naive} HR model for the $\mathrm{C_6H_{12}}$ foam and $\mathrm{C_6H_{12}Cu_{0.394}}$ foams, respectively. Our 1D modification to HR is introduced in the solid black and red curves. The Full 2D simple model is presented in the dashed curves. (b) The drive temperature $T_{D}$ is presented the green curve, the resulting $T_{S}$ for the 1D model for the pure $\mathrm{C_6H_{12}}$ foam is introduced in the black curve, when the 2D model is introduced in the dashed curve.}  
\label{fig:XvsTTWOP}
\end{figure}

In Fig.~\ref{fig:XvsTTWOP}(b) the resulting 1D (solid-blue) and 2D (dashed-black) surface temperatures $T_S(t)$, are presented. It can be seen that\, as expected, the temperatures used for the model are  significantly lower than the drive temperature $T_{D}$ (solid green), as was shown above in the Back et al. experiment (see Fig.~\ref{fig:XvsTBackPOP}(a) and the relevant discussion). We note again that the $T_{S}$ is estimated rigorously and not as a scaling factor, as was the case in previous works. These results show good agreement concerning the heat front breakout time (Fig.~\ref{fig:XvsTTWOP}(a)), whilst both the 1D and the 2D models yield close agreement to the experimental data (the experimental data lies between the 1D and the 2D models).
Specifically, the models predict (as do the full simulations) the difference in breakout time between the lengths that were tested ($300\mu$m-$400\mu$m). The variance between the pure and the doped foams is also shown in the analytic model.

\subsection{Back et al. $\mathrm{SiO_2}$ and $\mathrm{Ta_2O_5}$ high energy experiment}
\label{Back_high_exp}
These experiments were carried at the OMEGA-60 facility in 2000, and were then the most detailed supersonic heat wave experiments to date~\cite{Back2000,Landen}. The radiative drive temperature was shown previously in Fig.~\ref{fig:XvsTBackPOP} and reached the maximal temperature of $\approx190$eV. The experiment was carried out using two different foams, $50\mathrm{mg/cc}$ $\mathrm{SiO_2}$ and $40\mathrm{mg/cc}$ $\mathrm{Ta_2O_5}$ in several lengths (0.25-1.25mm). The samples were fabricated inside a cylinder, $1.6$mm diameter of different lengths, and were coated with $25\mu$m thick gold. The $\mathrm{Ta_2O_5}$ foam has a higher $Z$ than the $\mathrm{SiO_2}$ foam, and thus is more opaque due to its smaller Rosseland mean free path. However, the $\mathrm{SiO_2}$, has a larger heat capacity. Hence, the heat wave propagation is similar in both foams.

A comparison between the experimental results, the simple model and the simulations, presented and discussed in Sec.~\ref{sec:analytic} and Sec.~\ref{sec:Simulations} VI, for the $\mathrm{SiO_2}$ version of this experiment, demonstrates the importance of the 2D effects (energy wall loss and wall ablation) in this experimental setup.

\begin{figure}[htbp!]
\centering 
(a)
\includegraphics[width=7.5cm]{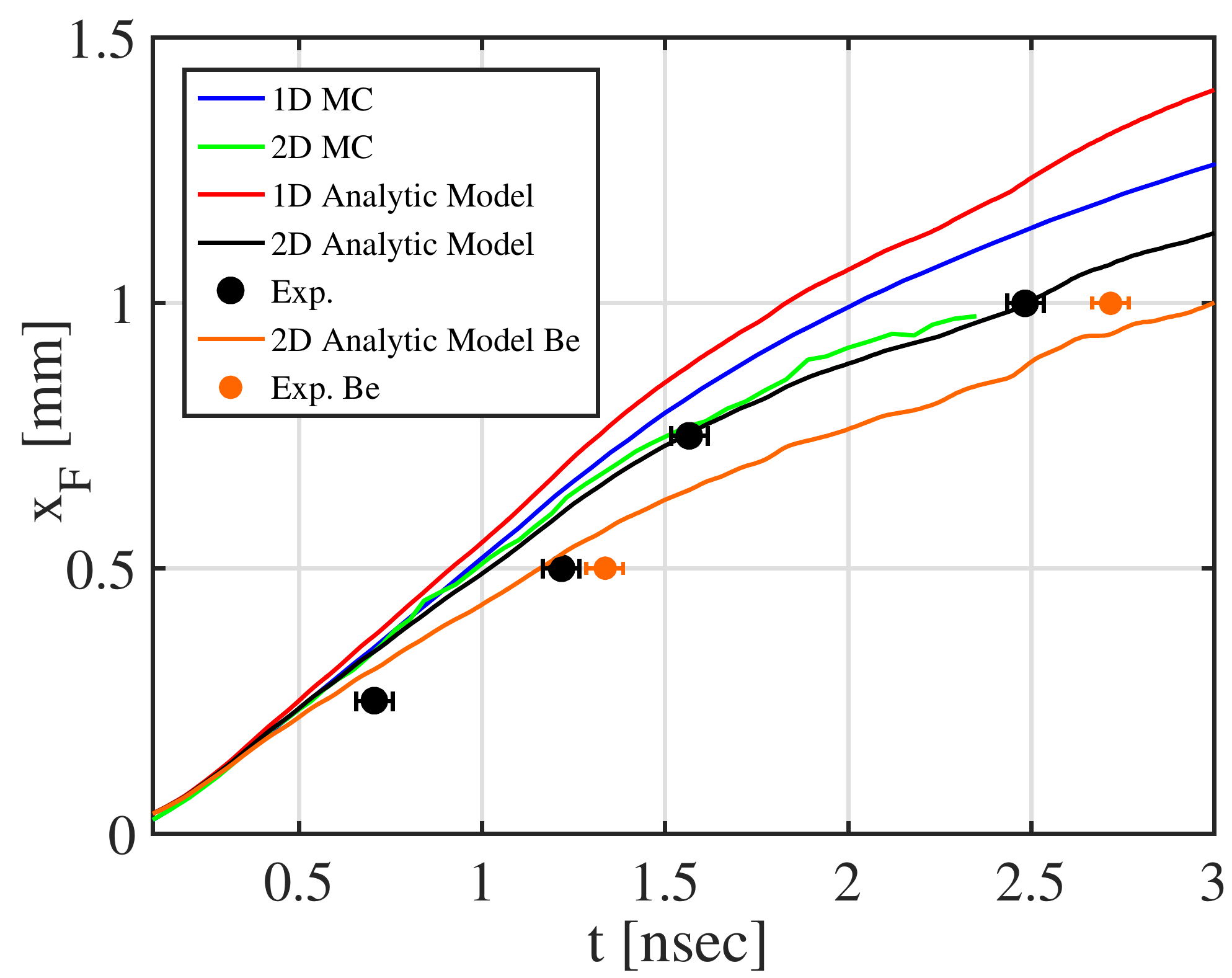}
(b)
\includegraphics[width=7.5cm]{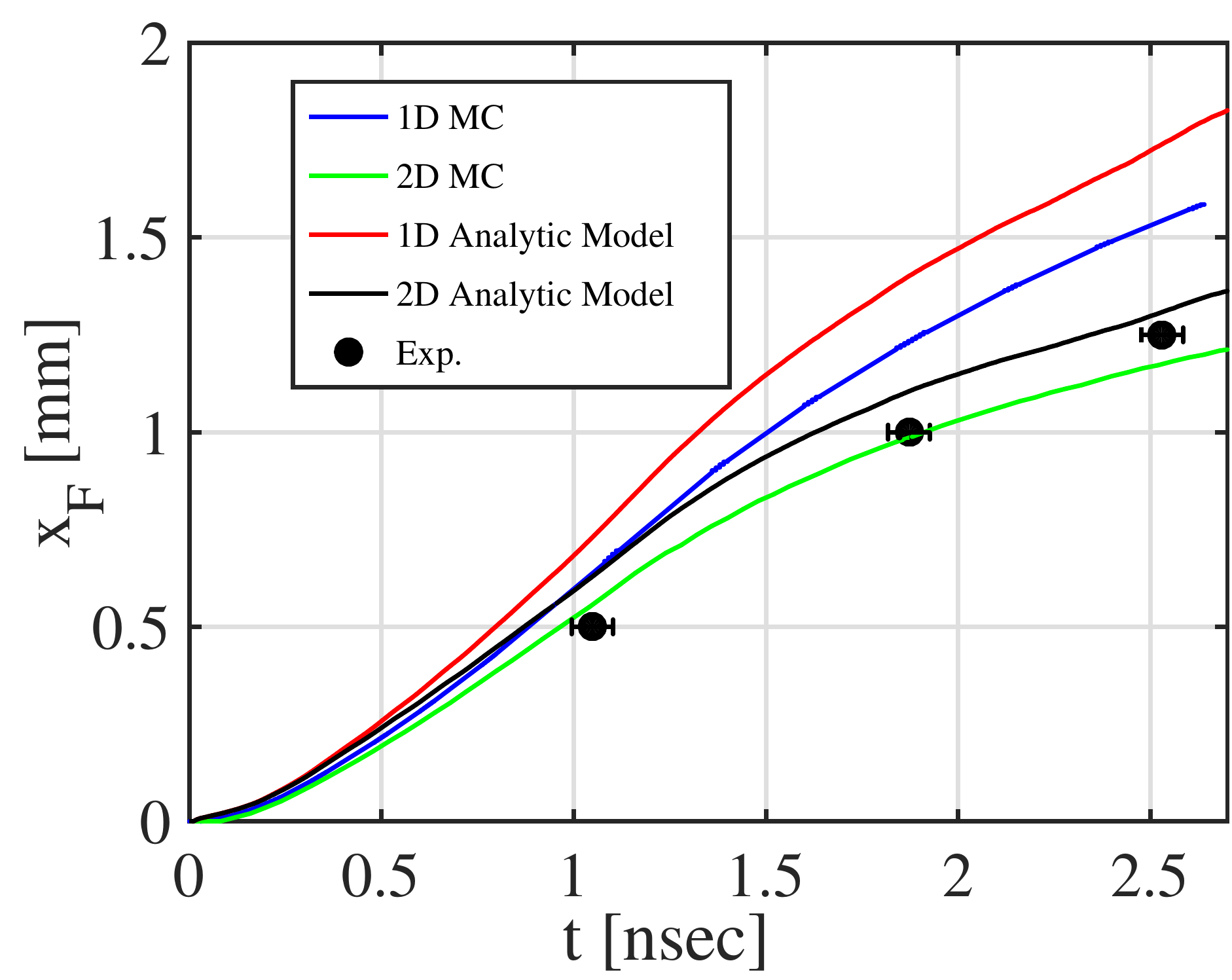}
\caption{The heat wave front $x_F(t)$ as a function of time for: (a) Back et al. high energy $Ta_2O5$ experiment and (b) the high energy $\mathrm{SiO_2}$ experiment~\cite{Back2000}. The experimental results are shown in the black circles, represent the time flux reached {at} half of its maximal intensity in the specific foam length. The red and black curves are for the 1D and 2D simple models, and the blue (1D) and green (2D) are for the IMC simulations. The simple model shows very good agreement with the simulations and the experimental results, when the dimensional effects can be also seen nicely. The orange circles in (a) are for $Ta_2O_5$ foam data, coated with a Beryllium sleeve instead of gold. The orange curve is the 2D analytic model for that experiment.}  
\label{fig:BACK2}
\end{figure}

The flux radiate that leaks from the edge of the foam was measured as a function of time for different lengths as in Fig.~\ref{fig:FLUXBack} for the $\mathrm{SiO}_2$ foams, showing good agreement between IMC simulations and experimental results, especially for the heat front breakout times (when flux reached half of its maximal intensity). The $\mathrm{Ta}_2\mathrm{O}_5$ foams show similar agreement. In Fig.~\ref{fig:BACK2}(a) we present the heat front position as a function of the time for the $\mathrm{Ta_2O_5}$ experiments, and in Fig.~\ref{fig:BACK2}(b) for the $\mathrm{SiO_2}$ experiments. First, the comparison between the 1D and 2D simulations and simple models shows that 2D effects in these experiments are dominant, and must be taken into account for the theoretical reproduction. Second, the full 2D simple model (black curves) has a high level of agreement with the experimental data. Finally, and as expected, the 2D IMC simulations (green curves) yield the best agreement with the experiment.

For the $\mathrm{Ta_2O_5}$ foam, an additional experiment was carried out, replacing the optically-thick gold sleeve with an optically-thin Beryllium sleeve~\cite{Back_BE,Landen,Hurricane2006}. The full experiment has not yet been published, but the reports available show that the heat front breakout times (with 0.5mm and 1mm foam lengths) are $10\%$ later, compared to the gold data (orange circles in Fig.~\ref{fig:BACK2}(a))~\cite{Landen}. This is due to the increased leakage of energy through the optically thin medium. The heat wave in the Be is almost supersonic (see Sec.~\ref{sec:analytic}) and energy leakage can be computed (see Eq.~18(b) in~\cite{Shussman2015}). We observe that the simple analytic model (orange curve) predicts the difference between the Be tube and the gold tube, in the 0.5mm foam it yields $10\%$ delay, while in the larger foam it is somewhat overestimates the time.
  
\subsection{Back et al. $\mathrm{SiO_2}$ low energy experiment}
\label{Back_low_exp}
An earlier, similar experiment of the Back et al.~$\mathrm{SiO_2}$ high energy experiment that was reported previously, took the form of a low energy analogy using $\mathrm{SiO_2}$ foam~\cite{BackPRL}.
This experiment was also carried out using the OMEGA-60 laser. The experiment was carried out at an extremely low density, $\rho=10\mathrm{mg/cc}$ $\mathrm{SiO_2}$, in low drive temperatures,  having $T_D^{\mathrm{max}}\approx85$eV. The laser pulse length was fairly long, $\sim{10}$ nsec. As for the high energy experiment, the samples were fabricated as cylinders of 1.6mm in diameter, in different lengths (0.5, 1 and 1.5mm), and were coated with $25\mu$m of gold.

\begin{figure}[htbp!]
\centering 
(a)
\includegraphics[width=7.5cm]{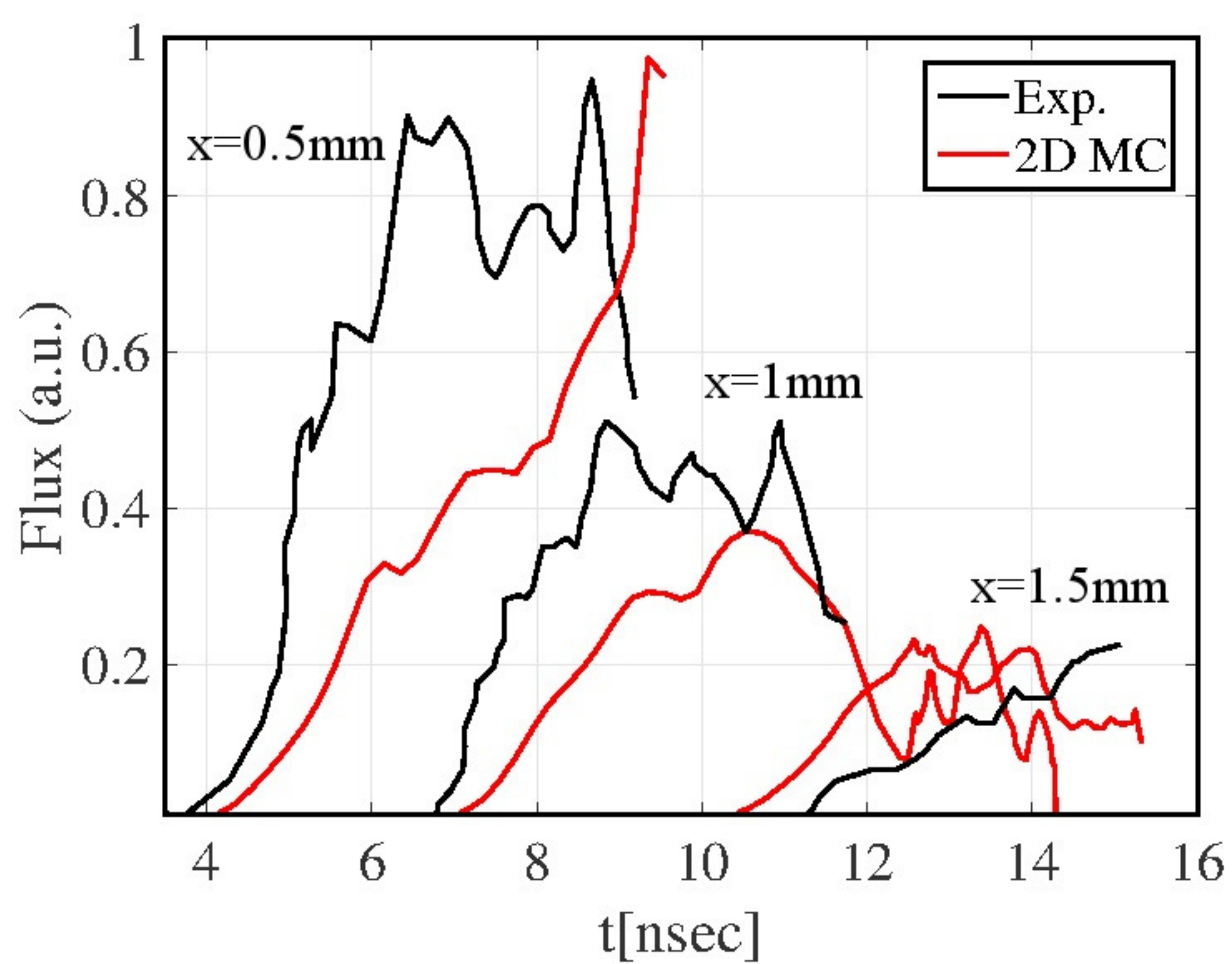}
(b)
\includegraphics[width=7.5cm]{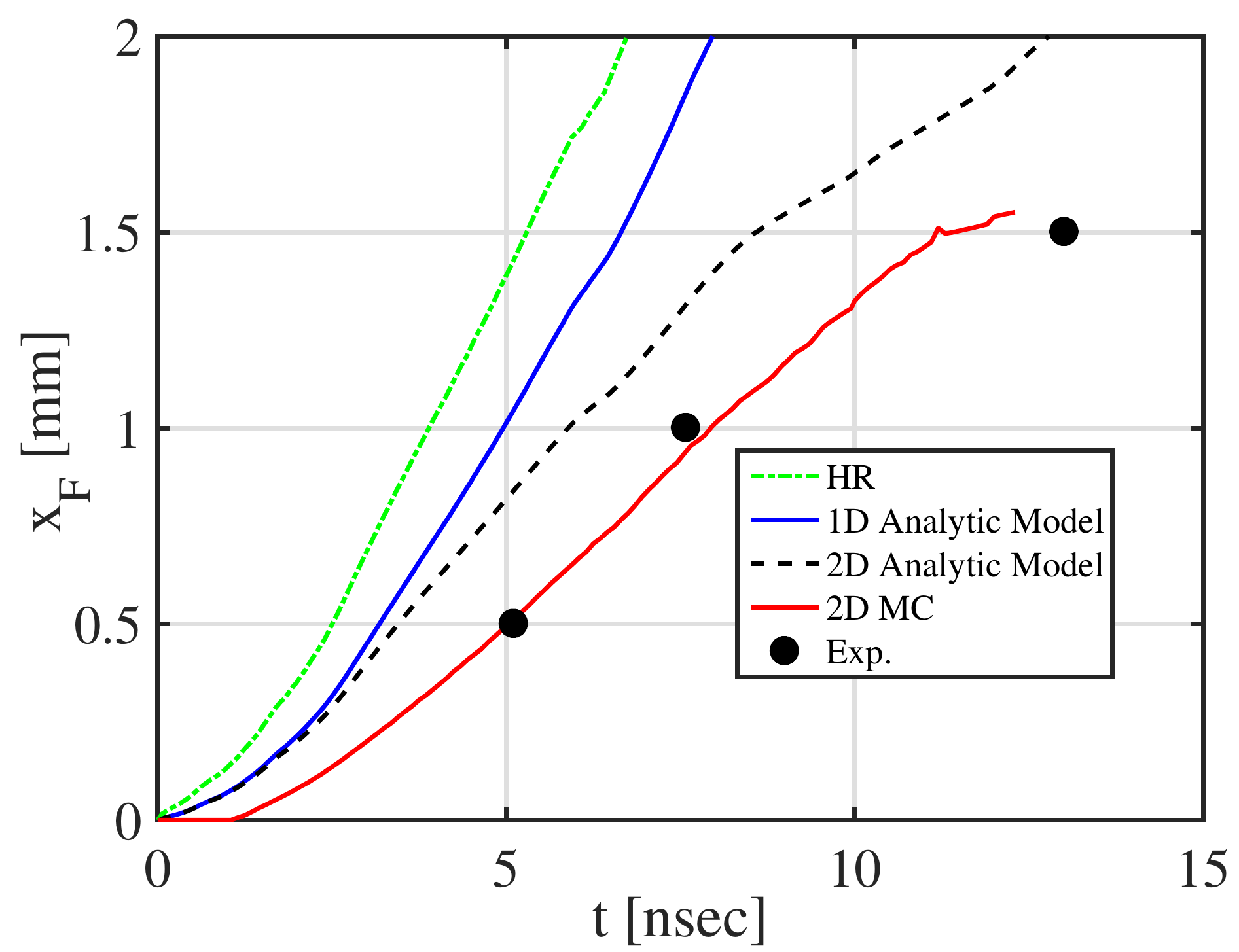}
\caption{(a) The radiated flux that leaks from the edge of the sample as a function of time for the low energy Back et al. $\mathrm{SiO_2}$ experiment~\cite{BackPRL}. The experimental data is in the black curves while IMC simulations are in red. (b) The heat wave front $x_F(t)$ as a function of time for different simple models in this experiment. HR prediction is in the green curve, while the corrected 1D simple model is in the blue curve. The full 2D simple model is in the black-dashed curve, and the full 2D IMC simulation is in the red curve. The experimental data is presented in the black circles.}  
\label{fig:BACK3}

\end{figure}

Since this experiment was conducted in a different thermodynamic regime (low temperature, low density), a specific new set of parameters was established for the semi-analytic model (see table~\ref{table:2} in the appendix).
A full comparison between the experimental flux that leaks from the edge of the foam and full 2D IMC simulations as a function of time is shown in Fig.~\ref{fig:BACK3}(a). Good agreement is evident again between the data and the 2D simulations for the breakout times. The agreement between the data and the simulations for the shape of the signal is only fair, although still within the typical agreement achieved in previous studies~\cite{BackPRL,back_china}. In Fig.~\ref{fig:BACK3}(b) we present a comparison of the heat front position data and the position predicted by the simple model (both 1D and 2D) and the simulations. For simplicity, the breakout time in the simulations was taken as when $T_m=25$eV, which is close to the time the flux reaches half of is maximum (the experimental breakout time) and the analytic model heat front definition. Note that the heat wave breakout time is not very sensitive to the exact value of $T_m$, within the range of $T_m=25\pm 5$eV (its maximal change is less than $0.4$ nsec). We can see that the 2D simple model (black-dashed curve) yields good agreement, especially in comparison to the 1D predictions (green and blue curves). This result implies that the 2D effects in these experiments are extremely important, due to the extremely long laser pulses used. The 2D simulations (red curve) yield very good results compared to the experiments, validating the macroscopic solution obtained. 

\subsection{The Moore et al. experiments (the Pleiades experiments)}
\label{Moore_exp}
To date, the most advanced and detailed experiments reported were published in several studies in 2015-2016 and are known as the Pleiades experiments~\cite{Moore2013,Moore2015,Guymer,Fryer2016}. These experiments were conducted in the high power NIF facility and the drive temperature had reached $\approx300$eV and shown in~\cite{Moore2015}. The experiment was carried out with two different foams, $\mathrm{C_8H_7Cl}$ (The Cl plays a major role in determining the foam opacity) and $\mathrm{SiO_2}$ in different densities (of about $100\mathrm{mg/cc}$). The physical packages in the experiments were all cylindrical, 2.8 mm long, and 2 mm in diameter, and enclosed in a $\mathrm{25\mu{m}}$ thick Au tube.

Two main diagnostics were used in the experiments. First is the Dante (an array of X-ray diodes), for measuring the radiative flux on the foam back side. A detailed analysis of the results from this diagnostic was presented in~\cite{Guymer,Fryer2016}, showing a non-negligible gap between the the experimental results and the theoretical predictions. This is because this experiment was designed in such a way that calculating the arrival time of the heat front to the end of the foam (2.8mm), is very sensitive to correct opacity and EOS~\cite{MoorePresentation}. However, unlike the Back experiments, the position of the heat front as function of time in this measurement has only one data point, concerning $x_F(t)$. This is because the length of the foam in these experiments was constant. 

\begin{figure}[htbp!]
\centering 
(a)
\includegraphics[width=7.5cm]{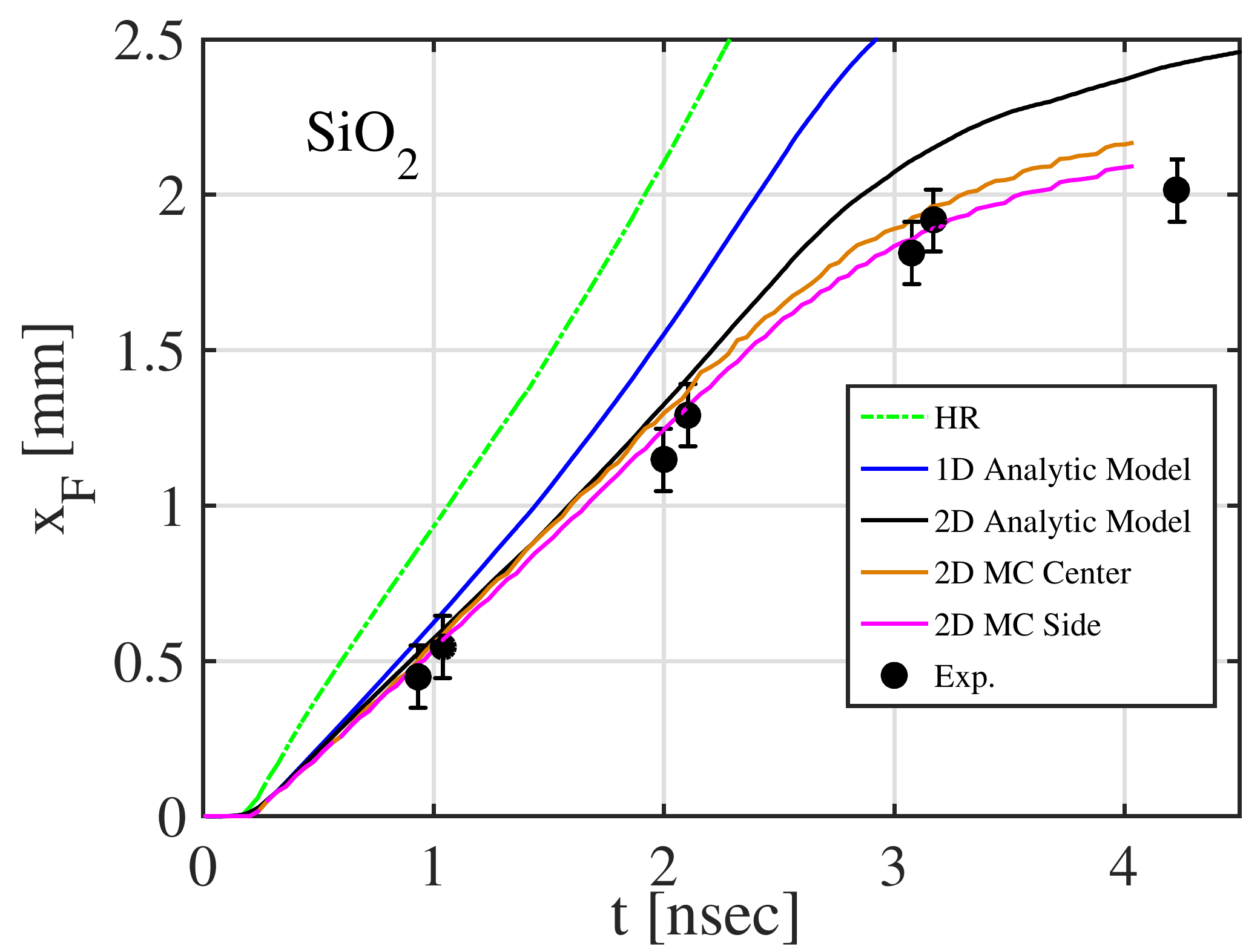}
(b)
\includegraphics[width=7.5cm]{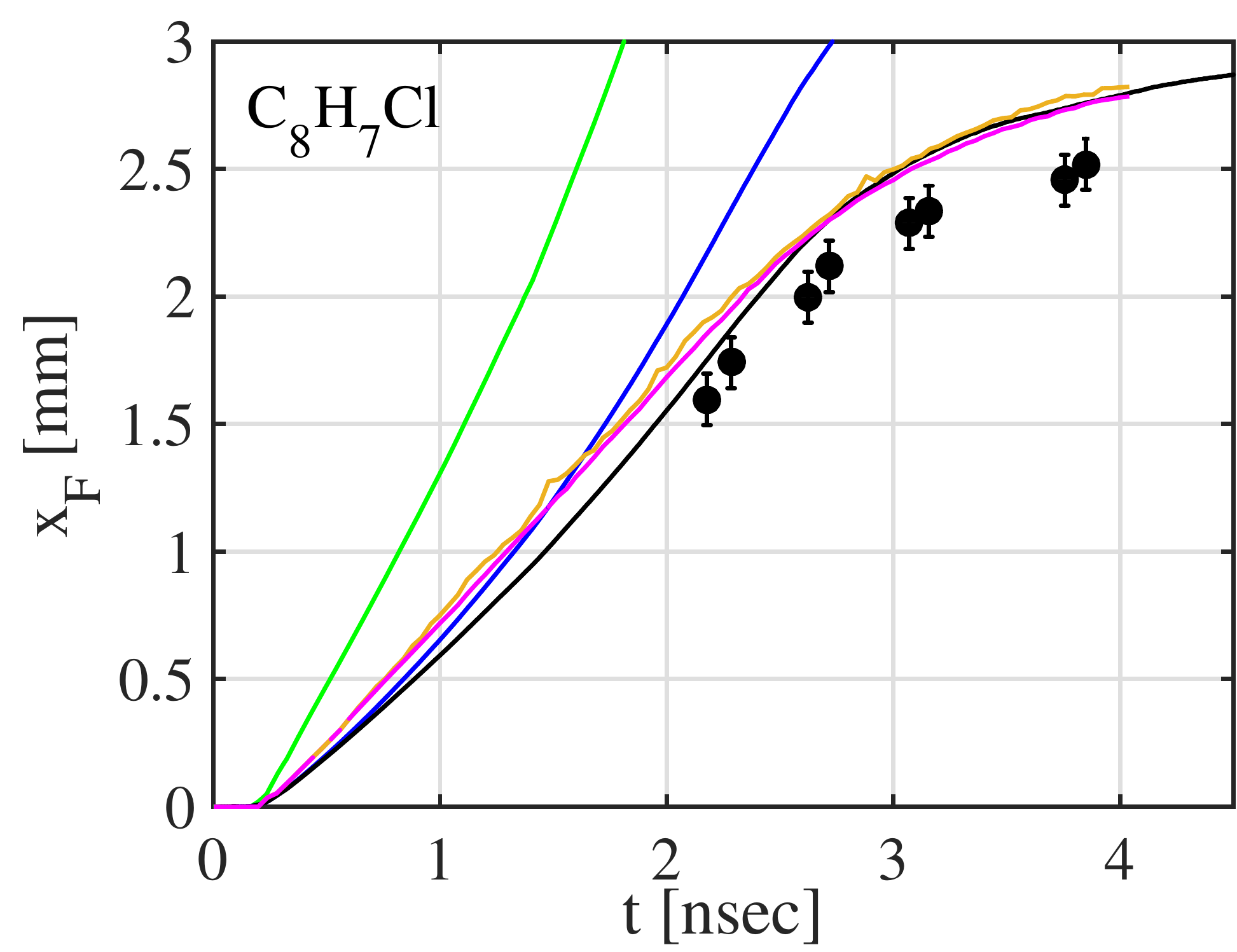}
\caption{The heat wave front position, $x_F(t)$, as a function of time for the different model versions and simulations for (a) $\mathrm{SiO_2}$ foam and (b) $\mathrm{C_8H_7Cl}$ foam~\cite{Moore2015}. The green curves are the
naive HR model, and the blue curves are our corrected 1D semi-analytic mode. The orange and magenta curves are the IMC simulations in the center and near the edge of the foams respectively. Our full 2D simple model is in the black curves showing good agreements with the IMC simulations in the center, and the experimental data, especially in the $\mathrm{C_8H_7Cl}$ foam.}  
\label{fig:MOORE}
\end{figure}

For tracking the heat front Eulerian position, $x_F(t)$, a second diagnostic was used, with the help of a window located in the side of the sample. This diagnostic constitutes the main interest of our own study. This measurement was carried out for two samples, $112\mathrm{mg/cc}$ of $\mathrm{SiO_2}$ and $114\mathrm{mg/cc}$ of $\mathrm{C_8H_7Cl}$ as shown in Fig.~\ref{fig:MOORE}(a),  and Fig.~\ref{fig:MOORE}(b). Several snapshots of the Moore et al. experiments using the 2D IMC simulations have been presented in Fig.~\ref{fig:MooreMap} (Sec.~\ref{sec:Simulations}), showing the importance of 2D effects in this experiment.

As can be seen in Fig.~\ref{fig:MOORE}, for both foams the naive HR model (using $T_S(t)=T_D(t)$, the green curves) extremely over estimates the front propagation velocity. Therefore Moore et al. used an {\em ad hoc} reduction factor of $0.71$,  for the drive temperature profile, $T_s$, similar to Massen et al. (who used a factor 0.8 of the heat wave velocity, see also Sec.~\ref{Massen experiments}). The 1D corrected model (blue curves), also over-estimates the heat front propagation velocity. However, the full 2D simple model (black curves) exhibits good agreement for both foams.

The full IMC radiative hydrodynamics simulations agree well with the experimental measurements. Two different IMC curves are presented in Fig.~\ref{fig:MOORE} for each foam. The orange curves represent the heat wave propagation on the center axis of the tube, and the magenta curves are near the side of the sample. We can see that for the $\mathrm{SiO_2}$ foam, a deviation between the two curves of simulations exists, when the side-simulation yields a slightly better accuracy. In the $\mathrm{C_8H_7Cl}$ foam, the difference between the two simulations is smaller. Since the simple model predicts the propagation in the center of the foam (due to its 1D basic feature), the model results in the $\mathrm{SiO_2}$ foam are a little poorer than in the $\mathrm{C_8H_7Cl}$ foam. However, in both foams the model results are in reasonable agreement with the simulation and the experimental data.

It should be noted that at late times, $\approx4$ nsec, the difference between the experimental data and the simulations increases. We attribute this to the rapid increase in heat front position sensitivity to opacity and EOS, when the heat wave reaches the edge of the foam at 2.8mm ~\cite{MoorePresentation}.

\subsection{Keiter \& Rosen et al. experiments}
\label{Keiter_exp}
As discussed above, the Xu et al. experiment~\cite{Dopped1,Dopped2} (Sec.~\ref{Xu experiments}), was a primary reported attempt to show the effect of doping with heavier metal (i.e. copper) on the heat front propagation. Keiter et al.~\cite{Rosen2007,Keiter2008} presented a series of experiments with $\mathrm{12\%}$ gold (high-Z) doped $\mathrm{C_{15}H_{20}O_6}$. Two different gold particle sizes were tested, checking the opacity model for atomically-mixed and/or finite-sized mix of gold particles.

The experiments were carried out in the OMEGA-60 facility, with a maximal radiation drive temperature of $\approx210$eV. All samples were made of $\mathrm{C_{15}H_{20}O_6}$ (pure or Au doped) with 62.5-65mg/cc, having a cylindrical shape with in 0.8mm diameter and in 1.2mm length, and were coated with gold. In the case of doping with large enough gold particles ($\approx5\mu$m or bigger), photons can flow around the gold particles, so the macroscopic opacity of the material is similar to pure foam's opacity. However, for small enough gold particles ($\approx{0.2}-1\mu$m), the doped foam converged to the atomic mix limit, and the Rosseland mean free path decreased rapidly as the doping level increase~\cite{Rosen2007,Keiter2008}.
\begin{figure}[htbp!]
\centering 
\includegraphics[width=7.5cm]{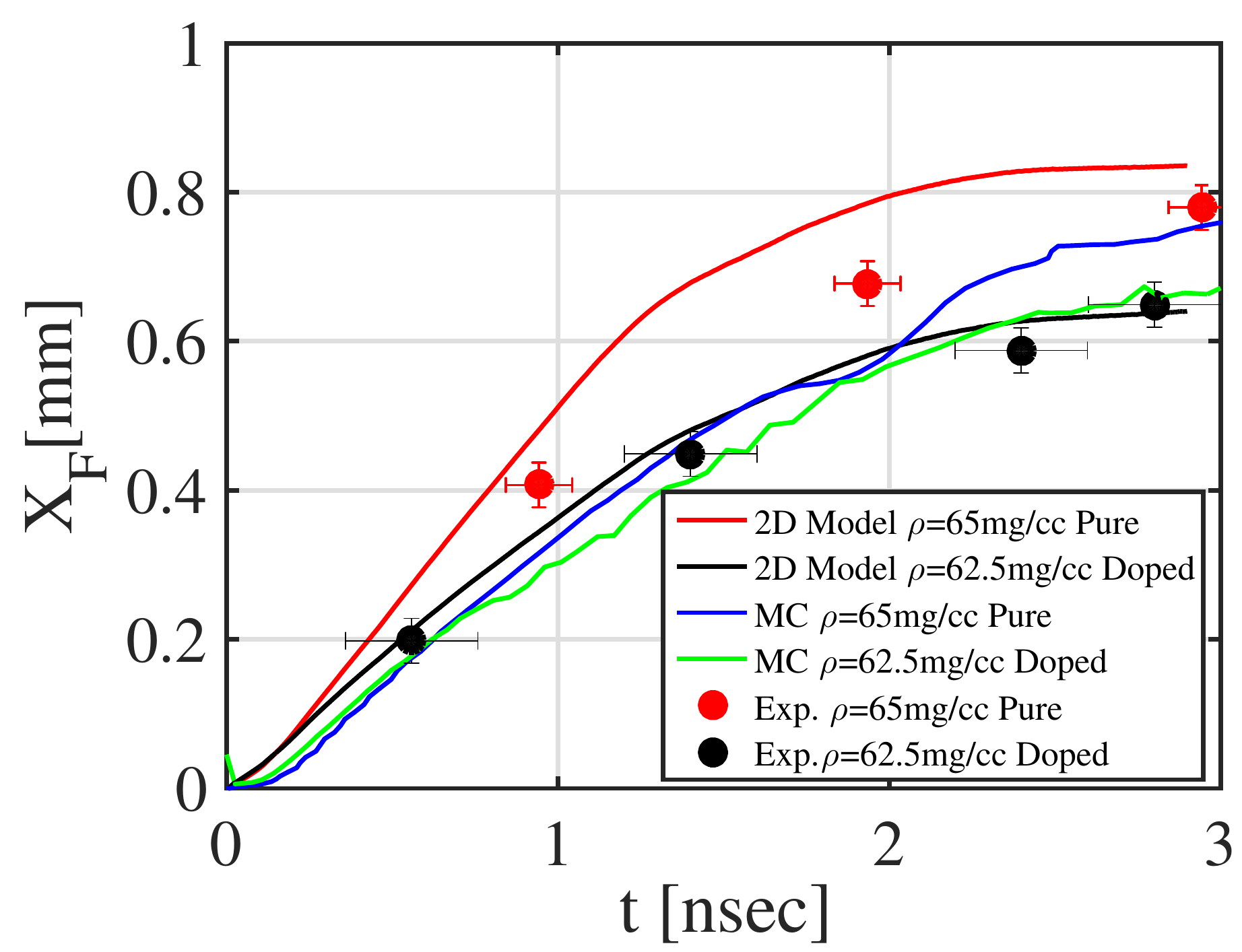}
\caption{The heat wave front $x_F(t)$ as a function for the Keiter et al. experiment~\cite{Keiter2008}. The red circles represent the experimental results for the pure foam, $\mathrm{C_{15}H_{20}O_6}$ with $\rho=65\mathrm{mg/cc}$. The black circles present the doped foam results with $12\%$ gold by mass, $\mathrm{C_{15}H_{20}O_{6}Au_{0.172}}$ with a little less density $\rho=62.5\mathrm{mg/cc}$. The red and the black curves are for the 2D pure and doped simple models. The blue (pure) and the green (doped) represent the 2D IMC simulations.}  
\label{fig:keiter}
\end{figure}

In these experiment the heat wave front position was measured by using a small window in the gold coating the foam, tracking the self emission of the hot foam. The heat wave front was defined as the place where the radiative flux reaches $40\%$ of its maximal value (matches to $T(x)\approx0.75T_{\mathrm{max}}$). In this case, we use Eq.~\ref{Henyey} with a Henyey-like profile (discussed in detail in Sec.~\ref{henyey_correction}). We note that in this experiment, 2D effects are extremely important and have to be take into account by using a 2D model.

The difference between pure versus doped foams can be clearly seen in Fig.~\ref{fig:keiter}. The red and black circles are the experimental data obtained for pure ($\mathrm{C_{15}H_{20}O_6}$ with $\rho=65\mathrm{mg/cc}$) and doped ($\mathrm{C_{15}H_{20}O_{6}Au_{0.172}}$ with $\rho=62.5\mathrm{mg/cc}$) foams, respectively (taken from~\cite{Keiter2008}). The 2D model prediction is given in the red curve for the pure foam, and in the black curve for the doped foam. Good agreement is evident between experimental data and the model prediction. We note that the 1D model (without 2D corrections) significantly deviates from the experimental data. The 2D IMC simulations (blue and green curves for the pure and doped foams, respectively), agree very well with the experimental results.

\subsection{Ji-Yan et al. experiment}
\label{C8H8_exp}
This experiment was performed on the SG-II laser, with a maximal radiation drive temperature of $\sim175$eV. The sample was a {\em bare} (without coating) plastic cylinder ($\mathrm{C_8H_8}$), with a density of $160\mathrm{mg/cc}$, $0.2$mm in diameter, and $0.3$mm long. The heat wave propagation was diagnosed by an XRSC that measured the self emission from the plastic perpendicular to the heat wave propagating. Since this foam is bare, the 2D model uses the radiating flux from the plastic to the surrounding vacuum, using Eq.~\ref{ELeak}. In addition, the heat wave front was defined as the place where the radiative flux reaches $50\%$ of the highest flux. Hence, we modify the heat front due to Eq.~\ref{Henyey} with $f=0.5$.

\begin{figure}[htbp!]
\centering 
\includegraphics[width=7.5cm]{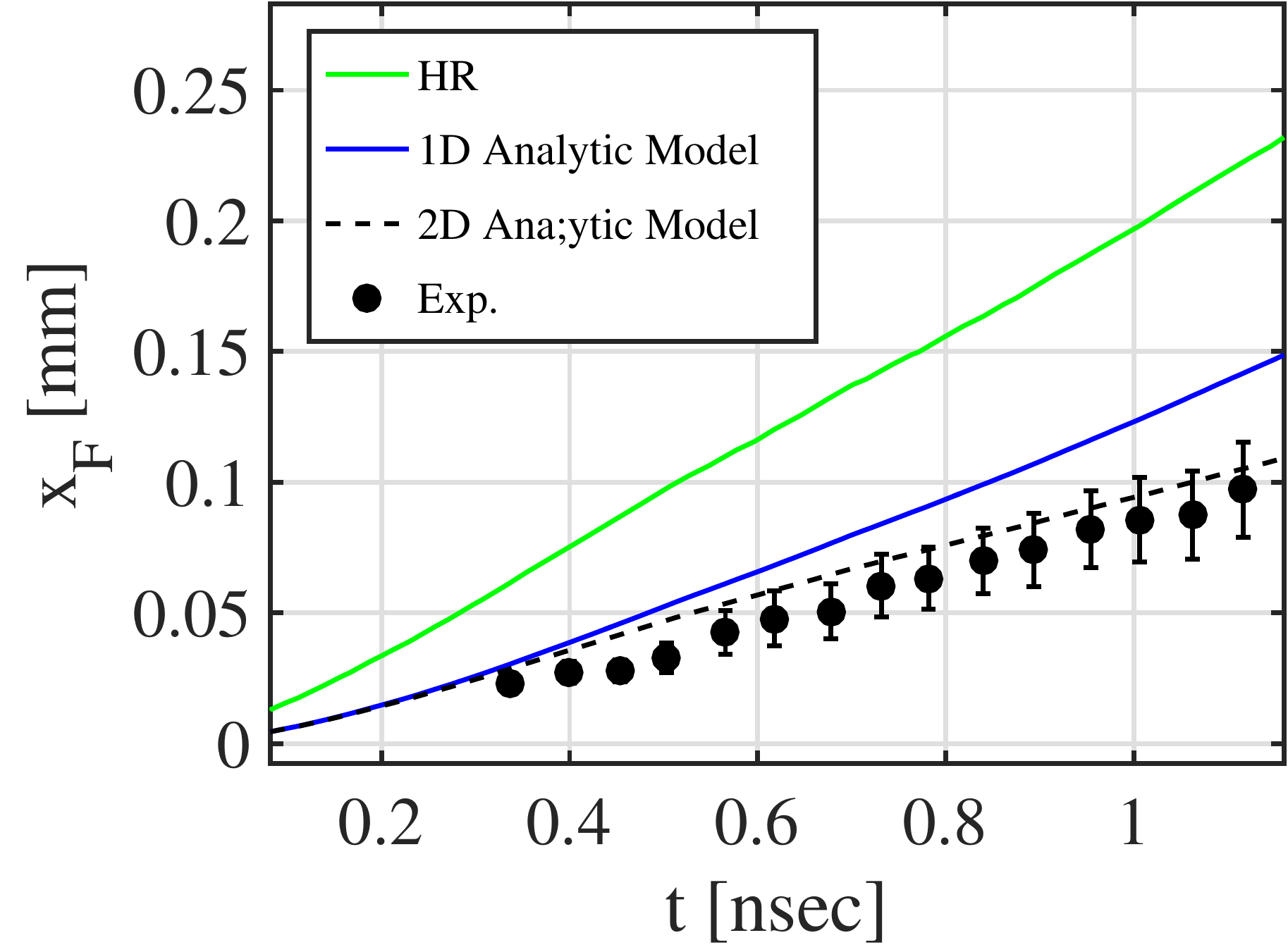}
\caption{The heat wave front $x_F(t)$ as a function of time for the Ji-Yan experiment~\cite{C8H8}. The experimental results from the $\mathrm{C_8H_8}$ foam are shown in the black circles. The different models results are shown in the green (naive HR), blue (1D corrected) and dashed-black (full 2D model) curves.}  
\label{fig:JiYan}
\end{figure}

The experimental data is shown by the black circles in Fig.~\ref{fig:JiYan}, while the different simple models are shown in the solid curves. We can see that the full 2D simple model again yields results that are very close to the experimental data, while 1D is too fast. This is due to the significant leakage of energy to the vacuum.

\section{Summary}
In the present work, a simple semi-analytic model for the radiative heat wave propagation in low density foams was presented and validated against a variety of experimental data, from different experiments carried out over the past three decades. 
The experimental results were also used to validate 2D numerical simulations using IMC for the radiative processes. Although the experimental setups that were examined employed different foams, densities and compounds, doping, dimensions,
etc., the simple semi-analytic model reproduces the experimental data very well, with differences of less than $10\%$. The model is based on the Hammer Rosen model, with the following modifications and improvements: 1. a correct boundary condition for the source temperature, taking into account the physical package - hohlraum surface; 2. matching of the front position definition to the experimental definition; 3. considering the energy loss to the walls; 4. the effect of wall ablation, including the 2D inlet narrowing of the
tube cross section and making the foam denser that slows the heat wave propagation. These simple building blocks allow the separation of variables, showing which physical phenomena dominates the experiment. 

IMC simulations showed very good agreement with the experimental data, matching not only in the heat front breakout times, but also in the full flux temporal profile. The simulations were also used to validate the model presented, allowing a better understanding of the governing processes and system dynamics. The good agreement between the model, experimental data and the 2D full simulations implies that the model correctly captures the physics dominating the problem. The fact that the same modeling applies to all the different experimental setups makes it possible, for the first time, to have all different experiments on one common ground.

\appendix
\section{Numerical Parameters for Material's Opacity and EOS}
\label{Appendix}
Using the exact opacity, CRSTA tables~\cite{Kurz2012,Kurz2013} and SESAME tables~\cite{SESSAME}, or QEOS~\cite{QEOS} EOS tables, we have fitted for every material its numerical parameters in the relevant experimental regime (by mean of temperatures and densities), for use in the different simple semi-analytic models. The parameters for the foams are above the double-line, and for the coats are beneath the double-line. The parameters for Au were taken from~\cite{HammerRosen}.
\begin{table}
\begin{center}
\begin{tabular}{||c | c | c | c | c | c | c | c ||} 
\hline
Experiment Name & Foam & $g [g/cm^2]$ & $f [MJ]$ & $\alpha$ & $\beta$ & $\lambda$ & $\mu$ \\[0.5ex] 
\hline
Massen  &  $\mathrm{C_{11}H_{16}Pb_{0.3852}}$ &  $1/3200$ & $10.17$ & $1.57$ & $1.2$ & $0.1$ & $0$ \\
\hline
Xu pure & $\mathrm{C_6H_{12}}$   &  $1/3926.6$ & $12.27$ & $2.98$ & $1$ & $0.95$ & $0.04$  \\
\hline
Xu with copper & $\mathrm{C_6H_{12}Cu_{0.394}}$   &  $1/7692.9$ & $8.13$ & $3.44$ & $1.1$ & $0.67$ & $0.07$  \\
\hline
Back, Moore  & $\mathrm{SiO_2}$  & $1/9175$ & $8.77$ & $3.53$ & $1.1$ & $0.75$ & $0.09$  \\
\hline
Back  &  $\mathrm{Ta_2O_5}$ & $1/8433.3$ & $4.78$ & $1.78$ & $1.37$ & $0.24$ & $0.12$  \\
\hline 
Back low energy & $\mathrm{SiO_2}$  & $1/9652$ & $8.4$ & $2.0$ & $1.23$ & $0.61$ & $0.1$  \\
\hline
Moore  & $\mathrm{C_8H_7Cl}$ & $1/24466$ & $14.47$ & $5.7$ & $0.96$ & $0.72$ & $0.04$  \\
\hline
Keiter Pure & $\mathrm{C_{15}H_{20}O_{6}}$ & $1/26549$ & $11.54$ & $5.29$ & $0.94$ & $0.95$ & $0.038$  \\
\hline
Keiter with Gold & $\mathrm{C_{15}H_{20}O_{6}Au_{0.172}}$ & $1/4760$ & $9.81$ & $2.5$ & $1.04$ & $0.35$ & $0.06$  \\
\hline
Ji-Yan & $\mathrm{C_8H_8}$  & $1/2818.1$ & $21.17$ & $2.79$ & $1.06$ & $0.81$ & $0.06$  \\
\hline
\hline
 & Au  & $1/7200$ & $3.4$ & $1.5$ & $1.6$ & $0.2$ & $0.14$ \\
\hline
 & Be  & $1/402.8$ & $8.81$ & $4.89$ & $1.09$ & $0.67$ & $0.07$  \\
\hline
\end{tabular}
\end{center}
\caption{All the materials parameters that were used in this paper in the simple semi-analytic models. The parameters are fitted to exact opacity, CRSTA tables~\cite{Kurz2012,Kurz2013} and SESAME tables (when appears)~\cite{SESSAME}, or QEOS~\cite{QEOS} tables in the experimental relevant regime. The parameters for foams are above the double line, and for the coats are beneath it.}
\label{table:2}
\end{table}

\begin{acknowledgments}
We acknowledge the support of the PAZY Foundation under Grant \textnumero~61139927. The authors thank Roee Kirschenzweig for using an IMC code for radiative problems. Special thanks to Mordecai (Mordy) Rosen from LLNL for the valuable discussions regarding the three different radiation temperatures and the wall-albedo.
\end{acknowledgments}

\setlength{\baselineskip}{12pt}

\end{document}